%% file: main.tex
\documentclass[sigconf, nonacm]{acmart}

\newcommand\vldbdoi{XX.XX/XXX.XX}

\newcommand\vldbavailabilityurl{https://github.com/Magolor/AgentHeaven}

\usepackage{times}
\usepackage{latexsym}
\usepackage{amsmath}
\usepackage{amsfonts}
\usepackage{amsthm}
\usepackage{bm}
\usepackage{url}
\usepackage{listings}
\usepackage{enumitem}
\usepackage{appendix}
\usepackage{tabularx}
\usepackage{multirow}
\usepackage{multicol}
\usepackage{enumitem}
\usepackage{booktabs}
\usepackage{siunitx}
\usepackage{makecell}
\usepackage{textcomp}
\usepackage{balance}
\usepackage{xcolor}
\usepackage{xspace}
\usepackage{setspace}
\setlist[itemize]{leftmargin=*, itemsep=0pt, topsep=2pt, parsep=0pt}

\definecolor{vldbBlue}{HTML}{2E4053}
\definecolor{vldbGreen}{HTML}{1E8449}
\definecolor{vldbRed}{HTML}{A93226}
\definecolor{vldbPurple}{HTML}{884EA0}
\definecolor{vldbOrange}{HTML}{D35400}
\definecolor{vldbGray}{HTML}{7F8C8D}
\definecolor{vldbBackground}{HTML}{F8F9F9}

\PassOptionsToPackage{noend}{algpseudocode}
\usepackage{algpseudocode}

\usepackage[lined,boxed,vlined,ruled,linesnumbered]{algorithm2e}
\definecolor{codegreen}{rgb}{0,0.6,0}
\definecolor{codegray}{rgb}{0.5,0.5,0.5}
\definecolor{codepurple}{rgb}{0.58,0,0.82}
\definecolor{backcolour}{rgb}{0.95,0.95,0.92}
\lstdefinestyle{codestyle}{
    backgroundcolor=\color{vldbBackground}, 
    commentstyle=\color{codegreen},         
    keywordstyle=\color{vldbBlue},          
    numberstyle=\tiny\color{vldbGray},      
    stringstyle=\color{codepurple},         
    basicstyle=\ttfamily\footnotesize,      
    breakatwhitespace=false,                
    breaklines=true,                        
    captionpos=b,                           
    keepspaces=true,                        
    numbers=left,                           
    numbersep=5pt,                          
    showspaces=false,                       
    showstringspaces=false,                 
    showtabs=false,                         
    tabsize=4,                              
    frame=single,                           
    rulecolor=\color{black!30},             
    title=\lstname                          
}
\theoremstyle{definition}

\usepackage[T1]{fontenc}

\usepackage[utf8]{inputenc}

\usepackage{microtype}

\usepackage{graphicx}

\begin{document}

\input{src/title}
\author{Zui Chen}\authornotemark[1]
\affiliation{%
  \institution{Huawei Company}
  \state{Shanghai}
  \country{China}
}
\email{chenzui1@huawei.com}

\author{Han Li}\authornotemark[1]
\affiliation{%
  \institution{Cornell University}
  \state{New York}
  \country{USA}
}
\email{hl2595@cornell.edu}

\author{Xinhao Zhang}\authornotemark[2]
\author{Xiaoyu Chen}\authornotemark[2]
\author{Chunyin Dong}\authornotemark[2]
\affiliation{%
  \institution{Huawei Company}
  \state{Shanghai}
  \country{China}
}

\author{Yifeng Wang}
\author{Xin Cai}
\author{Su Zhang}
\author{Ziqi Li}
\author{Chi Ding}
\author{Jinxu Li}
\affiliation{%
  \institution{Huawei Company}
  \state{Shanghai}
  \country{China}
}

\author{Shuai Wang}
\author{Dousheng Zhao}
\author{Sanhai Gao}
\affiliation{%
  \institution{Huawei Company}
  \state{Shanghai}
  \country{China}
}

\author{Guangyi Liu}
\authornotemark[1]\authornotemark[3]
\affiliation{%
  \institution{Huawei Company}
  \state{Shanghai}
  \country{China}
}
\email{liuguangyi5@huawei.com}

\thanks{$\ast$ These authors contributed equally to this work (co-first authors).}
\thanks{$\dagger$ These authors contributed equally to this work (co-second authors).}
\thanks{$\ddagger$ Corresponding author.}

\input{src/abstract}

\maketitle



\input{src/intro}
\input{src/related}
\input{src/method}
\input{src/benchmark}
\input{src/exp}
\input{src/limit}
\input{src/conclusion}

\clearpage
\bibliographystyle{ACM-Reference-Format}
\bibliography{bib/benchmark,bib/llm,bib/nl2sql,bib/kd,bib/agent,bib/rag}

\balance

\input{src/appendix}

\end{document}

%% file: src/title.tex
\newcommand{\sys}{\textsc{RubikSQL}\xspace}
\newcommand{\syss}{\textsc{RubikSQL-Lite}\xspace}
\newcommand{\sysm}{\textsc{RubikSQL}\xspace}
\newcommand{\sysl}{\textsc{RubikSQL++}\xspace}
\newcommand{\bench}{\textsc{RubikBench}\xspace}
\newcommand{\benchI}{\textsc{RubikBench0.9}\xspace}
\newcommand{\ukfI}{\textsc{UKF1.0}\xspace}
\newcommand{\ukf}{\textsc{UKF}\xspace}
\newcommand{\nlsql}{\textsc{NL2SQL}\xspace}

\title{\sys: Lifelong Learning Agentic Knowledge Base as an Industrial NL2SQL System}

%% file: src/abstract.tex
\label{sec:abstract}
\begin{abstract}
We present \sys, a novel \nlsql system designed to address key challenges in real-world enterprise-level \nlsql, such as implicit intents and domain-specific terminology. \sys frames \nlsql as a \emph{lifelong learning} task, demanding both \emph{Knowledge Base (KB) maintenance} and SQL generation. \sys systematically builds and refines its KB through techniques including database profiling, structured information extraction, agentic rule mining, and Chain-of-Thought (CoT)-enhanced SQL profiling. \sys then employs a \emph{multi-agent workflow} to leverage this curated KB, generating accurate SQLs. \sys achieves \emph{SOTA} performance on both the KaggleDBQA and BIRD Mini-Dev datasets. Finally, we release the \bench benchmark, a new benchmark specifically designed to capture vital traits of \emph{industrial \nlsql scenarios}, providing a valuable resource for future research.\footnote{Code \& Dataset: \texttt{URL\_TO\_BE\_RELEASED\_SOON}.}

\end{abstract}


%% file: src/intro.tex
\section{Introduction}
\label{sec:intro}
\begin{sloppypar}

\nlsql, translating natural language user queries into executable SQLs, has long been a valuable tool for simplifying data analytics. The recent advent of Large Language Models (LLMs) has proven to be a paradigm shift, with leaderboards of all major benchmarks like BIRD~\cite{BIRD} and Spider~\cite{spider} now primarily led by LLMs or fine-tuned LMs.

Nevertheless, significant challenges remain for real-world enterprise \nlsql applications. Such scenarios require the seamless integration of domain knowledge and user profiles during SQL generation. Financial data analytics is a prime example of this scenario. Queries often involve rich professional terminology and computation metrics, and originate from a diverse range of users within an organization. This makes it a highly valuable and representative case study for enterprise \nlsql.

Specifically, we concluded several vital challenges for real-world applications that require attention for practical utility:

\begin{itemize}
    \item \textbf{Implicit Intent}: While a standard and unambiguous query might be ``What is the YoY of revenue from product XXX sales in May 2025?'', real-world queries from financial analysts and management staff are often simplified and informal, such as ``XXX revenue last month?''. The system must infer the implicit intent behind these queries. For example, it should automatically identify YoY as the default metric, rather than the raw revenue amount only. Additionally, many entities in the database must be omitted or modified to align with the user's intent, which is often unstated. When creating a visualization of regional sales trends, the system must autonomously omit headquarters' data to prevent its outlier value from distorting the overall view. Similarly, when analyzing sales channels, it should group semantically similar channels like ``retail stores'' and ``kiosks'' to provide more meaningful insights without explicit user instruction.
    \item \textbf{Private Domain Knowledge}: This includes professional terminology, abbreviations, aliases of product names, temporary customer groups, self-defined financial metrics or formulas, differences in computation (e.g., financial vs. calendar year, currency exchanges), etc. While some \nlsql benchmarks present these settings as ``oracle knowledge'' or ``hints''~\cite{BIRD, KaggleDBQA} as one- or two-line notes, the rules for using different metrics can be too complex to summarize concisely in reality. For example, the same metric ``YoY'' can have different default behaviors for each table and different computation methods for different periods due to company data update and expiration policies, as there are separate columns for new and historical data.
    \item \textbf{Wide Table Schema}: Although the complexities in NL2SQL benchmarks often lie in table discovery and joins, another aspect of challenges to real-world financial \nlsql is the complexity of the schema of individual wide tables. Databases with trillions of rows often choose to transpose certain enums into additional columns, or vice versa, to improve SQL execution or storage efficiency. For instance, on small tables, currency exchanges might be a column predicate (\texttt{exchange = `USD'}) for flexibility, yet on large tables, each metric with an exchange is a separate column (e.g., \texttt{total\_revenue\_USD}) to reduce predicate computations. This results in similar queries having completely different SQL structures on different tables.
    \item \textbf{Context Sensitivity}: Context sensitivity is a crucial challenge where identical NL queries can be interpreted as different SQLs based on their context. For instance, temporal sensitivity in financial databases can be very counterintuitive to LLMs. Suppose a financial database updates on the 10th of each month. A query about ``this month'' on May 10th, 2025 should normally refer to May (\texttt{``202505''}), but on May 9th, it should refer to April (\texttt{``202504''}) to avoid empty responses. Furthermore, metrics like YoY require precise temporal framing; if it's May 2025, a YoY query should compare data from \texttt{``202501''} to \texttt{``202505''} with the same period in 2024 (\texttt{``202401''} to \texttt{``202405''}), not the full previous year. Besides time, user context also matters. If asked by a European staff member about local finance, the system should use EUR as the currency in the generated SQL. These seemingly subtle contextual factors are critical for the user experiences of \nlsql systems.
\end{itemize}

\begin{figure*}
    \centering
    \includegraphics[width=0.95\linewidth]{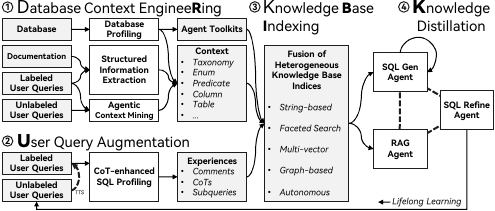}
    \caption{Overview of \sys, illustrating its four-stage knowledge-centric agentic workflow.}
    \Description{Overview of \sys, illustrating its four-stage knowledge-centric agentic workflow.}
    \label{fig:system}
\end{figure*}

In general, these challenges require a deep understanding of the database, its users, and their workloads, necessitating the accumulation of knowledge over time. We argue that an industrial \nlsql solution should not handle queries individually but rather operate on a \emph{long-term basis}. We believe that though generalization is useful for public benchmarks and cloud services, \emph{specialization} is much more critical for single-enterprise use cases. It is therefore essential to deploy a specialized, lifelong learning \nlsql system that \emph{resides within the database}. Such a system should grow and learn alongside its users, accumulating knowledge and adapting to evolving analytical trends, and transform the acquired knowledge into an \emph{AI-ready} representation.

However, by comparing real-world scenarios with frontier research \cite{CHASE,XiYanSQL,CSC,contextual,alphasql,Arctic,ODIS,ZeroNL2SQL,R1}, benchmark designs~\cite{BIRD,spider,KaggleDBQA,spider2,birdinteract}, and surveys~\cite{survey}, we observe that while nowaday \nlsql benchmarks tend to focus on evaluating the boundary of \nlsql techniques like table discovery, generalization across databases, multi-hop reasoning, they are not sufficient for assessing a method's deployability in actual enterprise scenarios.

Therefore, we propose \sys (Figure.~\ref{fig:system}), a systematic approach to enterprise-level \nlsql that integrates knowledge mining, maintenance and utilization as its core components, along with the \bench benchmark, a complex financial database with an anonymized schema from a real-world enterprise, to highlight the importance of lifelong learning in practical \nlsql applications.

The \sys system contains a suite of solutions for \emph{database context engineering} and \emph{user query augmentation} to construct and maintain an \emph{AI-ready} knowledge base. This knowledge base, coupled with an infrastructure providing indexing and search capabilities, supports a multi-agent workflow that adapts to user query workloads, continuously improving its performance through lifelong learning. The core concept of \sys is that such a well-maintained \emph{agentic knowledge base} is itself a model-agnostic solution to \nlsql.

In summary, our key contributions are:
\begin{itemize}
    \item \textbf{Unified Knowledge Format (\ukfI)}: We propose a knowledge representation format that acts as a semantic layer to decouple heterogeneous storage, indexing, and utilization. This enables LLMs and agents to uniformly access and maintain knowledge within the agentic KB.
    \item \textbf{Database Context Engineering}: We present a systematic solution for database context engineering, including \emph{database profiling}, \emph{structured information extraction}, and \emph{agentic context mining}. This process converts data from databases, documentation, and historical queries into \ukf, allowing \sys to adapt to specific scenarios.
    \item \textbf{Lifelong Learning}: We demonstrate the importance of lifelong learning in \nlsql, using techniques like \emph{CoT-enhanced SQL profiling} to extract experiences from user queries. These experiences are updated into the agentic KB, enabling the system to continuously self-improve.
    \item \textbf{Knowledge Base Indexing}: We propose an agent-driven approach that fuses multiple heterogeneous indices for schema linking and knowledge retrieval. Among the indices, we propose a novel \emph{LLM-augmented DAAC (Double Array Aho-Corasick automaton~\cite{DAAC}) index} that is both efficient and flexible.
    \item \textbf{SOTA Performance}: \sys achieves SOTA results on both BIRD Mini-Dev~\cite{BIRD} and KaggleDBQA~\cite{KaggleDBQA} (Table.~\ref{tab:bird_dev_results} and~\ref{tab:kaggledbqa_results}).
    \item \textbf{A Multi-Agent Workflow}: We explore a practical multi-agent workflow for utilizing the \ukf knowledge base, combining ICL (In-Context Learning) and CoT (Chain-of-Thoughts) for optimized performance. We also investigate the \emph{internalization} of knowledge into LMs via \emph{knowledge distillation}~\cite{cotd}, enabling smaller, faster, and more affordable enterprise \nlsql solutions.
    \item \textbf{\bench}: We introduce \bench to address the limitations of existing benchmarks for evaluating \nlsql systems in real-world enterprise scenarios. Compared to current benchmarks, \bench's three core contributions are its use of a realistic financial schema and workload; its focus on lifelong learning over a single database rather than generalization across multiple databases; and its inclusion of unique \emph{context-aware queries} that differentiate user profiles and preferences. Furthermore, we propose the \emph{Bipartite $F_\beta$-score} as a refined evaluation metric for \nlsql.
\end{itemize}

\end{sloppypar}

%% file: src/related.tex
\section{Related Work}
\label{sec:related}
\begin{sloppypar}

\subsection{LMs for \nlsql}
\label{sec:related-llms}

The application of Language Models (LMs) in \nlsql can be broadly categorized into 3 aspects:
\begin{itemize}
    \item \textbf{LLMs with Context Engineering}: This approach involves encoding user questions, database schemas, and few-shot in-context learning (ICL) examples into structured prompts. The prompts are then fed to LLMs or Reasoning LMs (RLMs) to generate valid SQL queries. Researches focus on optimizing prompt design for database profiles (e.g., M-Schema~\cite{XiYanSQL}), intelligent example selection/synthesis, and Test-Time Scaling (TTS) techniques~\cite{TTS}.
    \item \textbf{Fine-tuned LMs}: This category involves training one or a group of general or domain-specific LMs to directly produce SQL queries, often leveraging Supervised Fine-Tuning (SFT), LoRA~\cite{lora} or Reinforcement Learning (RL).
    \item \textbf{Agentic Pipelines}: These are multi-stage pipelines where various workers collaboratively generate a SQL query. The planning of such pipelines can range from fixed sequential steps to dynamic strategies, such as Monte Carlo Tree Search (MCTS)~\cite{alphasql} or the use of an LLM as an orchestrating agent. Workers (or tools) integrated into these pipelines may include retrieval mechanisms, SQL execution environments, or delegated LLMs for handling sub-tasks like schema linking, error correction, or rewriting.
\end{itemize}

Recent advances in RLMs~\cite{R1,o4-mini} have also led to increased focus on engineering reasoning traces, such as chain-of-thought (CoT)~\cite{step,cot}. In the context of \nlsql, this often translates to generating intermediate query plans or SQL drafts before the final SQL. Furthermore, knowledge distillation~\cite{cotd} has emerged as a novel LLM fine-tuning method, complementing fine-tuning techniques by adding CoTs produced by larger models as the supervision signals during training.

\subsection{Lifelong Learning and Agentic Memory}
\label{sec:related-memory}

To transition from one-off prompt engineering and model training, developing lifelong learning systems necessitates an agentic memory system. This usually involves a retrieval strategy and an update strategy. While many research efforts and projects have emerged~\cite{MemGPT,mem0,Zep,SecondMe,MemOS}, none have yet become dominant, and this remains an active field of research. A common trend in these agentic memory works points towards a heterogeneous memory architecture combining prompt engineering, hybrid Retrieval-Augmented Generation (RAG) systems, and parametric memory.

\subsection{Existing \nlsql Methods}
\label{sec:related-nl2sql}

Current leading \nlsql methods frequently combine multiple techniques from the categories above (Section.~\ref{sec:related-llms}). For instance, CHASE-SQL~\cite{CHASE} proposes a candidate-selection TTS framework. It employs query synthesis, ICL, and CoTs from multiple sources for candidate generation, and then a fine-tuned LM for selection. Similarly, XiYan-SQL~\cite{XiYanSQL} ensembles multiple ICL/SFT generators with additional fine-tuned LMs for selection/refinement, plus an LLM schema linking tool. Arctic~\cite{Arctic} and SQL-R1~\cite{SQL-R1}, primarily address the training aspects of \nlsql models. Their work involves synthesizing data for the SFT phase and designing effective reward functions for the RL phase.

The approaches above typically treat \nlsql as one-off tasks on static datasets rather than long-term deployed systems. AskData~\cite{AskData} was among the first to highlight the importance of query log analysis for real-world systems. Additionally, it examined various components of \nlsql tasks, including database profiling, schema linking, query plan processing, and feature extraction. Many recent works have also begun to recognize the importance of knowledge in \nlsql~\cite{TailorSQL,KAT-SQL}, yet addressing only single sources of knowledge fails to produce performance comparable to the well-optimized one-off methods above. And crucially, the absence of a unified knowledge representation and systematic update mechanism leaves this direction underexplored. \sys fills these gaps by proposing Unified Knowledge Format (\ukf) as structured knowledge representation to integrate lifelong learning and agentic memory (Section.~\ref{sec:related-memory}) into \nlsql, eventually achieving SOTA on both BIRD Mini-Dev~\cite{BIRD} and KaggleDBQA~\cite{KaggleDBQA} (Table.~\ref{tab:bird_dev_results} and~\ref{tab:kaggledbqa_results}), demonstrating the superiority of knowledge-centric \nlsql systems.

\end{sloppypar}

%% file: src/method.tex
\section{Overview}
\label{sec:method}
\begin{sloppypar}

\subsection{Problem Definition}
\label{sec:method-definition}

Unlike conventional approaches to \nlsql, we formulate real-world enterprise \nlsql as a {\it lifelong learning} task over a relatively stable database (where schemas remain largely consistent over time).

Let $\mathcal D = (\mathcal S, \mathcal V, \tilde{\mathcal K}^0)$ denote a stable database with schema $\mathcal S$, content $\mathcal V$, and initial, raw knowledge input $\tilde{\mathcal K}^0$ (encompassing schema semantics, terminologies, implicit intents, historical user queries, etc.). The system maintains a dynamic knowledge base $\mathcal K^t$ and responds to NL queries via two coupled components: a {\it KB update module} $\mathcal M_k$ and a {\it SQL gen module} $\mathcal M_s$:
\begin{align}
\Delta \mathcal K^t &= \mathcal M_k\left(q_t, \mathcal{S}, \mathcal V, \mathcal K^{t-1} \right)\label{eq:knowledge_update_module} \\
s_t &= \mathcal M_s\left(q_t, \mathcal S, \mathcal{K}^t \right) \label{eq:sql_generation_module}
\end{align}

Given a batch of queries with expected outcomes, denoted as $Q = \left\lbrace (q_t, o_t^\ast) \right\rbrace_{t=1}^T$, the objective is to maximize the {\it Execution Accuracy} (EX) of generated SQLs $\lbrace s_t \rbrace_{t=1}^T$:
\begin{equation}
\max_{\mathcal M_{s}, \mathcal M_{k}} \mathcal A = \max_{\mathcal M_{s}, \mathcal M_{k}} \frac{1}{T} \sum_{t=1}^T \mathbb{I}[\texttt{Exec}(s_t) = o_t^\ast]
\end{equation}

It is worth addressing that though the architectures of $\mathcal M_k$ and $\mathcal M_s$ ought to be generalizable, the KB $\mathcal K^t$ is \emph{instance-optimized}, evolving over user queries accumulated over time. The database schema $\mathcal S$ and domain knowledge set $\mathcal K^0$ are generally assumed to remain in-distribution, permitting only gradual, sparse updates. Once setup, the KB $\mathcal K$ is not aimed for transfer to other Out-Of-Distribution (OOD) databases. Nevertheless, OOD queries is supported.

It is also worth emphasizing the {\it transductive} nature of lifelong learning: for streaming queries, subsequent queries can leverage context from prior unlabeled queries; while for batched queries, each query may leverage knowledge from all unlabeled queries in of before this batch. 

\subsection{Overall Architecture}
\label{sec:method-architecture}

Under the aforementioned setting (Section.~\ref{sec:method-definition}), we propose {\textbf \sys}, a four-stage knowledge-centric agentic workflow comprising: Database Context Enginee{\textbf R}ing, {\textbf U}ser Query Augmentation, Knowledge {\textbf B}ase {\textbf I}ndexing, and {\textbf K}nowledge Distillation (Figure.~\ref{fig:system}). \sys also introduces the \emph{Unified Knowledge Format} (\ukf, Section.~\ref{sec:method-ukf}) as a semantic layer for representing knowledge to bridge the four stages, leading to an \emph{agentic knowledge base}.

\emph{Database context engineering} transforms the initial knowledge base $\tilde{\mathcal K}^0$ into $\mathcal K^0$, while \emph{user query augmentation} contributes to both the construction and update of the KB $\mathcal K^t$, functioning as the \textit{KB update module} $\mathcal M_k$. This module enables structured mining of context entries and experiences from multiple sources, including database schema, documentation, and historical user queries. After construction, the KB $\mathcal K^t$ has all its context entries and experiences stored in \ukf. Such a semantic layer decouples knowledge storage and mining from indexing and further utilization.

The KB organized in \ukf supports fusion of heterogeneous indices through a unified retrieval interface (Section.~\ref{sec:method-index}). Built upon this well-maintained, easily-accessible, AI-ready knowledge base, the \textit{SQL gen module} $\mathcal M_s$ employs a multi-agent workflow to generate SQL queries. This workflow includes a \emph{RAG Agent}, a \emph{SQL Gen Agent}, and a \emph{SQL Refine Agent} that interact with each other and with the KB $\mathcal K^t$ to produce robust, verified SQL queries. Technically, \sys combines In-Context Learning (ICL) and Chain-of-Thought (CoT) for SQL generation. Through \emph{knowledge distillation}~\cite{s1K,cotd}, \sys improves SQL generation effectiveness and efficiency via Supervised Fine-Tuning (SFT) over carefully curated NL-CoT-SQL training tuples. \sys further optimizes the accuracy-cost trade-off with cascading (e.g., using models of different sizes like 14B and 32B) and Test-Time Scaling (TTS) techniques (e.g., majority voting).

In the following sections, we first present \sys's knowledge representation format — Unified Knowledge Format 1.0 (Section.~\ref{sec:method-ukf}) — then elaborate on the \textit{KB update module} $\mathcal M_k$'s components in Section.~\ref{sec:method-kb}, and finally describe the \textit{SQL gen module} $\mathcal M_s$ in Section.~\ref{sec:method-use}.

\end{sloppypar}

\section{Knowledge Base Maintainance}
\label{sec:method-kb}
\begin{sloppypar}

\subsection{Unified Knowledge Format 1.0}
\label{sec:method-ukf}

\sys introduces \ukfI (Unified Knowledge Format 1.0) as a knowledge representation that systematically organizes information from heterogeneous sources such as database schemas, documentation, and historical user queries into a standardized structure.

Designed for convenient, automated, and long-term maintenance within enterprise systems, the \ukf serves as an underlying protocol that supports multiple storage backends (e.g., memory, file systems, databases) and diverse search indices. It also includes additional features like provenance tracking, version control, and knowledge base personalization.

For \ukfI's detailed definition, see Appendix~\ref{sec:appendix-ukf}. Briefly summarizing, \ukfI categorizes knowledge attributes into six core groups:
\begin{itemize}
    \item \textbf{Metadata}: Descriptive attributes (e.g., \texttt{name}, \texttt{type}, \texttt{version}) with both human-oriented and LLM-oriented descriptions.
    \item \textbf{Content}: Primary knowledge content that supports both natural language descriptions (\texttt{content}) and semi-structured data (\texttt{content\_resources}). For semi-structured data, the \texttt{content\_composer} mechanism enables dynamic serialization through configurable functions, allowing context-aware, language-aware, and model-aware prompt construction. For example, a natural language domain knowledge record describing how a specialized ``YoY'' metric should be computed in a company could be stored as \texttt{content}; a dictionary of column taxonomies that contains the functional dependency related to column could be \texttt{content\_resources}; prompts and serializers can be stored as \texttt{content\_composer}.
    \item \textbf{Provenance}: Source tracking attributes (e.g., \texttt{source}, \texttt{owner}, \texttt{creator}, \texttt{parents}, \texttt{auths}), enabling personalized KBs with privacy and authority controls.
    \item \textbf{Retrieval}: Optimization attributes for search indices (e.g., \texttt{tags}, \texttt{synonyms}) and filtering (\texttt{triggers}). For example, \texttt{tags} contain slot-value pairs for faceted search over \ukf instances, while \texttt{synonyms} enable string-based indexing (notably the LLM-augmented DAAC index; Section.~\ref{sec:method-index-str}). Search indices (Section.~\ref{sec:method-index}) handle coarse-grained retrieval, whereas filtering refines the final knowledge set on a per-instance level. The \texttt{priority} attribute resolves conflicts during knowledge selection.
    \item \textbf{Relationships}: A special part of the optimization attributes designed for graph-based indexing. These are knowledge graph tuples (\texttt{subject}, \texttt{predicate}, \texttt{object}, \texttt{metadata}), where predicates may reference other UKF instances and can also be UKF instances themselves.
    \item \textbf{Life-cycle}: Temporal management attributes (e.g., \texttt{timestamp}, \texttt{last\_verified}) enabling dynamic updates, lazy deletion, and version control.
\end{itemize}

\begin{figure}
    \centering
    \includegraphics[width=1.00\linewidth]{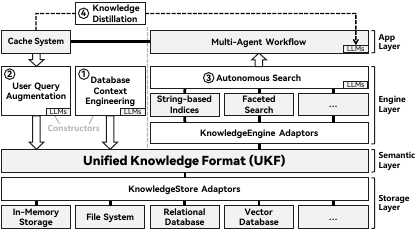}
    \caption{Layered architecture built upon \ukf, decoupling knowledge storage and mining from indexing and utilization.}
    \Description{Layered architecture built upon \ukf, decoupling knowledge storage and mining from indexing and utilization.}
    \label{fig:architecture}
\end{figure}

Besides definition, The \ukf also offers built-in functionalities to simplify the management and creation of new entries. For a specific scenario, templates inheriting the \ukf can be created conveniently by defining constructors in natural language and \texttt{content\_composer} (See Appendix~\ref{sec:appendix-ukf-def} for details about creating \ukf templates). Here are some useful general-purpose templates:

\begin{itemize}
    \item \textbf{Knowledge}: Free-form, short domain knowledge records, often supplied by humans or extracted via Structured Information Extraction (SIE, Section.~\ref{sec:method-mining-sie}).
    \item \textbf{Document}: Chunks of documentation used as sources for SIE.
    \item \textbf{Experience}: Historical user queries, both labeled and unlabeled. \sys provides built-in components that can monitor input-output pairs to accumulate experiences in \ukf. The accumulated Experiences can be easily converted into few-shot prompts and training data. A useful variant of these templates is LLM messages, which allows knowledge to be managed as SFT data. Another variant, specific to \nlsql, is NL-SQL pairs with enriched information (CoTs, etc.) obtained from SQL profiling.
\end{itemize}

Specifically for \nlsql applications, \sys implements the following specialized, built-in \ukf templates:
\begin{itemize}
    \item \textbf{Table}: Information about a table.
    \item \textbf{Column}: Information about a column.
    \item \textbf{Enum}: Information about a value in a column.
    \item \textbf{Taxonomy}/\textbf{Dependency}: The relationship between columns or values in a table (mainly functional dependencies).
    \item \textbf{Predicate}: Complex \texttt{WHERE} clause conditions.
    \item \textbf{Synonym}: Context-aware synonyms for other \ukf instances. For example, ``Lincoln'' can refer to a person, a band or a car under different context.
    \item \textbf{Indicator}/\textbf{Term}/\textbf{Metric}: Domain-specific formulas (e.g., \texttt{profit = revenue - cost}) and computation metrics (e.g., YoY, CAGR).
    \item \textbf{Special}: Special rules (e.g., temporal, data validity, domain knowledge, implicit intents).
\end{itemize}

As a logical knowledge representation protocol, \ukf is agnostic to its physical storage and indexing implementations (Figure.~\ref{fig:architecture}). Nevertheless, \sys contains built-in adapters that conveniently connect \ukf to in-memory storages, file systems, and databases.

With the base \ukf definition, built-in variants, and implementation-agnostic storage and indexing, \sys can swiftly convert a database and related documentation into an initial agentic KB $\mathcal K^0$ with minimal human effort (Section.~\ref{sec:method-mining}). Following the initial engineering, \sys continuously updates its KB as it accumulates an increasing number of user requests. The update include both \emph{agentic context mining} (Section.~\ref{sec:method-mining-agent}) and \emph{user query augmentation} (Section.~\ref{sec:method-aug}). This dynamic updating process leverages both vast amounts of unlabeled user queries and occasional labeled NL-SQL pairs through periodic updates, enabling the system to refine its context and improve performance over time.

\subsection{Database Context Engineering} 
\label{sec:method-mining}

Upon deployment, \sys first transforms the initial domain knowledge, denoted as $\tilde{\mathcal K}^0$, into executable agentic toolkits and structured \ukf instances through systematic \emph{database context engineering}.

Database context engineering in \sys is achieved by multiple agentic workflows, including \emph{Database Profiling} (Section.~\ref{sec:method-mining-dbprof}), \emph{Structured Information Extraction} (Section.~\ref{sec:method-mining-sie}), and \emph{Agentic Context Mining} (Section.~\ref{sec:method-mining-agent}). The system extracts knowledge from four primary sources: database schemas and content, documentation, labeled NL-SQL pairs, and unlabeled user NL queries.

\subsubsection{Database Profiling}
\label{sec:method-mining-dbprof}

Database profiling is a necessary component for adapting \nlsql solutions to specific databases, as proven in almost all \nlsql researches~\cite{MAC-SQL,XiYanSQL,AskData}. To minimize human efforts, \sys features a general-purpose database profiling for basic column and table statistics, and then an LLM-based descriptor to generate comprehensive tables and column descriptions.

Unlike traditional database profiling, \sys considers \emph{agentic toolkits} an extra form of database profile. To strengthen the agents' understanding of the database, \sys equips them with specialized database access toolkits, including \texttt{column\_info}, \texttt{table\_info}, \texttt{db\_info}, \texttt{execute\_sql}, \texttt{fuzzy\_column}, and \texttt{fuzzy\_enum}.

To support these tools, \sys first generates a general-purpose profile for columns, tables, and the database.

For column profiling, \sys first performs column type annotation using simple rules (numeric, categorical, text) and LLMs. For example, if a column is suspected to be of temporal type, an LLM is asked to generate a \texttt{strptime} format string to parse the column's values. \sys then constructs column profiles, which include general attributes (ID, name, description, PK/FK status, type, \texttt{NULL} count) and type-aware attributes:
\begin{itemize}
    \item For numerical columns: mean, min, 25/50/75 percentiles, max, and whether the data is integer or float.
    \item For categorical and textual columns: number of classes, most common values with corresponding counts, frequencies, accumulated frequencies, and max string length.
    \item For temporal columns: time format string, min time, and max time.
\end{itemize}
Then, missing column descriptions are automatically populated by an LLM. Table and database profiling follow a similar approach, capturing structural metadata while using an LLM to supplement missing descriptions.

For SQL execution, the \texttt{execute\_sql} tool not only returns the SQL query results in an LLM-friendly format but also automatically truncates returned results, preserving statistical information and errors to optimize context window usage.

For fuzzy search operations, \sys combines lexical and semantic matching. For lexical matching, since LLMs typically generate short keywords rather than long queries, \sys utilizes the \emph{asymmetric Jaccard containment} score computed over n-grams of tokenized values as the distance metric. The score is defined as $J = \vert Q \bigcap D \vert / \vert Q \vert$, where $Q$ represents the short query keywords and $D$ denotes the enum value. For semantic matching, \sys uses a general-purpose frozen embedding model (e.g., BGE~\cite{BGE}).

\subsubsection{Structured Information Extraction}
\label{sec:method-mining-sie}

While enums, columns, and tables can be obtained directly from a database schema, other context entries like domain-specific predicates and indicators require external input. Ideally, \sys can quickly adapt to a new scenario by directly importing \ukf instances. However, the difficulty of converting knowledge from unstructured sources into \ukf could result in extra manual effort. Therefore, performing automated structured information extraction (SIE) on existing enterprise documentation can significantly reduce the manual effort required to deploy \sys.

Since most benchmarks like BIRD~\cite{BIRD} and Spider~\cite{spider} do not provide documentation as a source, we rely on established open-source solutions for SIE. In real-world scenarios, due to the clear and straightforward format of \ukf, large language models (LLMs) orchestrated with existing SIE solutions~\cite{llamaextract,pz,langextract}, self-implemented SIE, or simply LLMs that support structured outputs, can all automatically parse natural language into \ukf instances with satisfactory precision using built-in \ukf adapters (Figure.~\ref{fig:architecture}). Nevertheless, just as we emphasize in \sys that knowledge updates and SQL generation should be an interleaving process, the same should apply to SIE when facing extraction tasks with complex domain knowledge. We leave the research of a more tightly integrated, self-improving SIE process as a future direction.

\subsubsection{Agentic Context Mining}
\label{sec:method-mining-agent}

User queries serve as another significant source of context entries in real-world enterprise \nlsql systems.

For labeled user queries (i.e., NL queries with ground-truth SQLs), linking phrases in an NL query to clauses in SQL is a relatively easy task for modern LLMs, which are capable of producing a structured JSON that can then be converted to \ukf. For example, from the correspondence between the user query ``What are the top 10 countries with the most number of running plants?'' and the SQL \texttt{SELECT "Country" FROM "nuclear\_power\_plants" WHERE "Status" = "Operational" GROUP BY "Country" ORDER BY COUNT("Name") DESC LIMIT 10}, LLMs can extract \text{Table}: \texttt{"nuclear\_power\_plants"} with synonym ``plants'', and \text{Enum}: \texttt{"Operational"} with synonym ``running''. To improve robustness, the system can first generate a SQL without looking at the ground-truth. If the generated SQL's execution result doesn't match the ground-truth's, this erroneous SQL can be included in the extraction prompt. In this way, LLM is instructed to compare the key differences between the correct and incorrect SQLs, helping it identify the specific predicates or conditions that led to the failure. This process results in more robust \ukf extraction.

Unlabeled queries pose greater challenges, often comparable in difficulty to the \nlsql task itself. Nevertheless, it is still worthwhile to perform agentic context mining on these queries because the process runs \emph{offline}, removing the need for low latency. This allows \sys to use larger LLMs, longer reasoning traces, and agents with more steps and tools. Furthermore, \sys can now adopt advanced techniques like Test-Time Scaling (TTS), including majority voting, reward or reranker models~\cite{XiYanSQL}, and Reflexion~\cite{Reflexion}.

Specifically, the context mining agent for unlabeled queries resembles the SQL generation process, leveraging both the toolkits described in Section.~\ref{sec:method-mining-dbprof} and the indexing methods from Section.~\ref{sec:method-index} to retrieve relevant information about the query. The key differences are that, first, instead of returning a SQL, the agent is only asked to retrieve relevant information, which reduces the complexity of SQL structure sketching and reasoning; second, the context mining agent can use more resources than the online SQL generation agent as explained above, and its results can be verified by comparing them with existing knowledge in the KB. The context mining agent then produces a structured JSON output, similar to the labeled case. Alternatively, the system can also generate pseudo-labels (SQLs) for unlabeled queries and perform AST (Abstract Syntax Tree) parsing based on the SQL, regardless of its correctness. The AST approach allows for a more elaborate examination of the extracted context, such as verifying whether a column or value truly exists.

The generated JSON context entries, after conversion to \ukf, are merged with existing context according to the following order of preference: (1) context entries with identical content and ID merge their synonyms; (2) entries that conflict with existing schema, labeled, or human-verified context entries are discarded; (3) all entries with identical IDs and conflicting contents are dropped; (4) the remaining entries are considered valid, but are inserted into the KB with a lower priority than schema, labeled, or human-verified entries.

\subsection{User Query Augmentation} 
\label{sec:method-aug}

As discussed in Section.~\ref{sec:method-ukf}, \textbf{Experience} represents a specialized type of knowledge: NL-SQL pairs that can serve as training data or few-shot examples. However, when LLMs are trained or prompted with raw NL-SQL pairs, especially those with long, complex SQL structures, they often fail to capture critical information. This can result in poor comprehension of the logical reasoning process and, consequently, subpar few-shot performance. To address this limitation, \sys introduces \emph{CoT-enhanced SQL Profiling}, which adds enriched context to raw NL-SQL pairs to help improve LLM and agent understanding.

\subsubsection{CoT-enhanced SQL Profiling}
\label{sec:method-aug-sql}

Mainstream LLMs typically acquire coding capabilities from pretraining on extensive human code repositories with rich commentary. In contrast, SQL statements in \nlsql datasets and real-world enterprise requests often lack comprehensive comments, as most are single-use or generated via BI dashboards. \sys addresses this gap by enhancing raw SQLs with comments, thereby transferring LLMs' comment reading capabilities to the SQL domain. This process enforces explicit reasoning by LLMs and helps clarify user intent. Furthermore, \sys expands NL-SQL pairs to NL-CoT-SQL pairs to enable knowledge distillation (Section.~\ref{sec:method-inf-data}).

The augmentation process begins with the original NL-SQL pair. First, query time is added. For temporal scenarios in real-world applications, query time is crucial for determining query predicates. For example, interpreting ``What's the gross profit this month?'' requires the current month. Without this temporal information, historical queries can become misleading examples.

Next, the schemas of the involved tables are obtained from executing the corresponding SQL. This helps clarify user intent and enables potential output format personalization. In many \nlsql datasets and real-world scenarios, a seemingly incorrect SQL (judged by exact matching) may actually be logically valid but simply fail to produce the exact columns the user desires. For example, when asked about ``revenue'', most LLMs would return a single revenue amount column by default. However, when the experiences are clearly marked with an output schema like (2024 amount, 2025 amount, YoY), it instructs the LLM to write a SQL that includes the $YoY$ information the user implicitly wants.

Then, important \ukf retrieved from the indices (Section.~\ref{sec:method-index}) are summarized and placed in the SQL header. For example, the definition of ``the most important customers'' should be based on noteworthy recent activities instead of overall scale. Notifying LLMs with such an explanation in the header can be very beneficial for its understanding of the SQL.

After collecting these attributes, \sys uses an LLM to generate an AI-friendly version of the SQL. The LLM receives the raw NL-SQL pair, the collected attributes, and the \ukf instances as input. It is instructed to perform two specific tasks: (1) produce an analysis section as the SQL's comment header, which contains all the aforementioned attributes, a concise description of the SQL draft in NL, and key corner cases (e.g., \texttt{NULL}, zero division); and (2) output the SQL with inline comments for each clause. For an example of a profiled SQL with commentary, see Appendix.~\ref{sec:appendix-sql}.

In addition to the NL-based SQL descriptions in the comment header, \sys leverages separate reasoning CoTs generated by modern RLMs to provide additional supervision signals. These reasoning CoTs are particularly useful for knowledge distillation (Section.~\ref{sec:method-inf}). Specifically, given an NL-SQL pair with extracted knowledge, a large RLM is instructed to generate a SQL query along with its CoT. If the generated SQL is correct, the experience is updated as an augmented NL-CoT-SQL tuple. If the generated SQL fails to produce the same result as the ground truth SQL, the RLM is instructed to modify its CoT with minimal adjustments, using the ground truth SQL as a reference. The modified CoT is then re-verified by checking whether continued inference from the updated CoT leads to a successful SQL generation.

\subsubsection{Query Synthesis}
\label{sec:method-aug-query}

Data synthesis is a widely used approach for modern LLM instruction tuning~\cite{self-instruct}. However, synthesized NL-SQL pairs often suffer from limited diversity~\cite{Arctic}, whether they are generated from a schema or through query paraphrasing.

Our key observation is that LLMs can generate higher-quality queries through a process of \emph{composition} and \emph{decomposition} rather than simple paraphrasing. Therefore, instead of prompting LLMs to diversify NL queries by generating queries of similar difficulty, \sys prompts them to generate either simplified versions of complex queries, or complicated versions of simple queries.

For simplified queries, \sys relies solely on LLM-generated NL and SQL decompositions, which is typically correct when the original ground-truth NL-SQL pair is provided. For example, given the query ``Which department, A or B, had the greatest increase in important customers last year?'', simplified queries can be generated, such as ``What was the increase in the number of important customers for department A last year?'' or ``What was the increase in the number of important customers for department B in January last year?''. Similarly, extracting CTEs from a complex SQL is a relatively easy task for modern LLMs.

For complicated versions, rather than running the entire SQL generation workflow, \sys prompts a standalone LLM to follow specific CTE templates (with a \texttt{WITH}, \texttt{UNION}, or \texttt{JOIN} structure) to perform SQL merging. This is often achieved using knowledge types like Indicator, Term, and Metric (Section.~\ref{sec:method-ukf}). For example, given the terms ``VAT'' and ``Discount'', and the query ``What is the VAT for order XXX?'', a simple composed question would be ``What are the VAT and Discount for order XXX?''. Given an indicator like ``profit = revenue - cost'', complex queries about profit can be created by combining simple CTEs that query revenue and cost. Similarly, a metric like ``YoY = \texttt{CASE WHEN <period\_start> = 0 THEN 0 ELSE (<period\_end>-<period\_start>) / ABS(<period\_start>)}'' can be added to almost any count-based indicator, converting it to a rate-based indicator after YoY. Combining predicates can be tricky and often leads to queries that do not make sense in the real world (e.g., an employee living in both region A and B). In this case, the database ontology — which contains taxonomies and dependencies (Section.~\ref{sec:method-ukf}) — is essential for the composition process, as it prevents the specification of conflicting predicates by only drawing random predicates from independent ontology groups (e.g., customers, regions, departments).

In addition to query synthesis, for benchmarks like BIRD \cite{BIRD}, the train, dev, and test sets use different databases, lacking in-distribution training data. Therefore, we also employ \emph{query transfer}. Given the original NL-SQL pair on the source dataset, we first extract knowledge related to the ground-truth SQL from the original database. Then, given the ontology, schema, and agent toolkits to access the target database, LLMs are instructed to generate new SQLs that have a similar structure to the original SQL but are executable on the target database. Finally, given the original NL-SQL pair with the new SQLs, corresponding NLs are generated through SQL2NL~\cite{AskData}. For selecting high-quality synthesized queries, we prioritize generated SQL with meaningful, non-empty execution results.

\end{sloppypar}

\section{Knowledge Base Utilization}
\label{sec:method-use}
\begin{sloppypar}

\subsection{Knowledge Base Indexing} 
\label{sec:method-index}

\sys provides flexible storage options for mined \ukf context entries and experiences, including in-memory key-value stores, database records, JSON files in a file system, or blob storage. Built on top of this storage layer, a separate, \textit{storage-independent} engine layer (index layer) retrieves \ukf using various methods (Figure.~\ref{fig:architecture}). \sys implements the following KB indices: string-based index, faceted search, multi-vector index, graph-based index, and autonomous search. Specifically, we propose the novel \emph{LLM-augmented DAAC index} as an efficient and practical implementation of the string-based index.

\subsubsection{String-based Index} 
\label{sec:method-index-str}
Apart from traditional techniques like LSH, edit distance, or embedding-based semantic similarity, \sys recommends exact pattern matching with LLM-augmented synonyms as the string-based index, namely the \emph{LLM-augmented DAAC index}. A small LM generates synonyms for \ukf instances; for example, it might suggest ``operational'' as a synonym for ``functioning''. These synonyms are then lemmatized and processed by a DAAC (Double Array Aho-Corasick automaton~\cite{DAAC}), which can find all matching synonyms for a query string $Q$ in linear time $O(\vert Q \vert + r)$, where $\vert Q \vert$ is the length of the query string and $r$ is the number of matches. Since the DAAC query time is independent of the number of patterns or the lengths of patterns, it is extremely efficient when matching millions of knowledge entries. Compared to a traditional AC automaton, DAAC provides a significant constant-factor speedup and dramatic memory savings.

While traditional DAAC-based lexical matching struggles with abbreviations and aliases, LLM-augmented synonyms mitigate this issue without introducing extra query complexity. In practice, we found that the DAAC approach with augmented synonyms achieves sufficient retrieval recall while significantly improving retrieval speed. Nevertheless, it is worth noting that the recall of the LLM-augmented DAAC can vary across different languages, as languages with fewer tenses, variations, and genders are better suited for synonym-based matching. To address this, we added a lemmatization layer to all query keywords and synonyms before feeding them into the DAAC.

\subsubsection{Faceted Search}
\label{sec:method-index-facet}
Faceted search in \sys is implemented using a dedicated SQL database that indexes key attributes of \ukf instances. The primary facet is \texttt{tags}, a set of slot-value pairs describing the context where a \ukf is applicable. For example, the predicate \texttt{WHERE "Status" = "Operational"} can be annotated with tags like \texttt{[ENUM=operational]}, \texttt{[COLUMN=status]}, and \texttt{[TABLE=nuclear\_power\_plants]}. Note that a single slot can hold multiple values.

To support efficient multi-faceted queries, including partial and wildcard matching (e.g., via the \texttt{LIKE} operator), we use a normalized schema. Each multi-value facet attribute is stored in a separate table and linked to its corresponding \ukf instance via a foreign key (id). Multi-value facet attributes include \texttt{tags}, \texttt{related}, \texttt{synonyms}, and \texttt{auths}. For other attributes, we build indices on retrieval, provenance, and metadata to accelerate query performance, leveraging the native index types supported by various SQL dialects.

\subsubsection{Multi-vector Index}
\label{sec:method-index-vector}
The multi-vector index in \sys supports indexing on user-designated embeddings or on embeddings of serialized \ukf instances via type-specific serializers. Each embedding type resides in a separate vector index, supporting both naive ANN (Approximate Nearest Neighbor) search and MMR (Maximal Marginal Relevance) search. Retrieved \ukf instances from multiple indices are deduplicated and merged via reranking or round-robin selection. In \sys, this index is primarily used for the \emph{Experience} knowledge type. We implement the following key serialization functions:
\begin{itemize}
    \item {\textbf{Query}}: The original user natural language query is used directly.
    \item {\textbf{Query Sketch}}: Entities such as columns, tables, and enums are replaced with type placeholders~\cite{AskData,MCS}. For example, for ``List the revenues of top 10 customers from North America.'', a basic query sketch could be ``List the [COLUMN] of top [VALUE] [COLUMN] from [ENUM].'', while an advanced taxonomy-aware sketch may yield: ``List the [REPORT\_ITEM as COLUMN] of top [VALUE] [CUSTOMER as COLUMN] from [REGION as ENUM].''
    \item {\textbf{Tags}}: A sorted, concatenated list of tags serves as the key. For the example query, this could be: \texttt{[COLUMN:customer]|[COLUMN:region]|[ENUM:North America]|[COLUMN:report\_item]|[ENUM:revenue]}. In practice, tags also include metadata such as query time, user identity, and inferred query types.
    \item {\textbf{SQL}}: An initially generated SQL query is used to find similar examples~\cite{ODIS}.
    \item {\textbf{Header}}: The comment header from an initially generated SQL serves as the search key.
    \item {\textbf{CoT}}: An initially generated chain-of-thought (CoT) rationale is used for retrieval.
\end{itemize}

\subsubsection{Graph-based Index}
\label{sec:method-index-graph}
For the graph-based index, \sys leverages the \texttt{related} tuples within \ukf definitions to construct a knowledge graph. This graph supports few-hop searches conditioned on relation types, such as linking predicates to specific enums, enums to columns, and dependencies to taxonomies.

\subsubsection{Autonomous Search}
\label{sec:method-index-auto}
Autonomous search enables agents to use all the index APIs to generate candidates. For instance, an LLM can specify keywords for a string-based search. Retrieved candidates are automatically filtered by evaluating their \texttt{triggers} to exclude those unsuitable for the current scenario. The LLM agent then selects the most relevant subset of retrieved \ukf instances or creates a natural language summary, effectively functioning as a Retrieval-Augmented Generation (RAG) agent.

\subsection{SQL Generation}
\label{sec:method-inf}

During inference, \sys's SQL gen module, $\mathcal{M}_s$, uses three agents to generate a SQL query: the \textit{RAG Agent} (Section.~\ref{sec:method-index-auto}), the \textit{SQL Gen Agent}, and the \textit{SQL Refine Agent}. Each agent has a distinct function:

\begin{itemize}
    \item The \textbf{RAG Agent} processes the user's query and uses knowledge base (KB) indices (Section.~\ref{sec:method-index}) and toolkits (Section.~\ref{sec:method-mining-dbprof}) to retrieve relevant \ukf instances. For deployment in query-intensive and resource-constrained environments, this agent can be configured as a static pipeline with pre-defined reranking and selection policies. For example, it could return all knowledge from the LLM-augmented DAAC index and the top-5 results from the round-robin multi-vector index. After retrieval, a one-off LLM is instructed to summarize these retrieved results.
    \item The \textbf{SQL Gen Agent} takes the user's query and the retrieved \ukf instances (with optional toolkits) as inputs to produce a CoT-enhanced SQL query with commentary information.
    \item The \textbf{SQL Refine Agent} receives the user's query, the SQL generated by the SQL Gen Agent, and related \ukf instances as inputs. It then uses a single SQL execution tool to verify the generated SQL. Based on the execution results or errors, it corrects and optimizes the SQL query.
\end{itemize}

By offloading agentic operations to the RAG and Refine Agents, the SQL Gen Agent can be simplified to a single, fine-tuned SQL Gen LM. This is achieved through knowledge distillation~\cite{cotd}. The outputs from the RAG Agent (NL + \ukf) are accumulated and paired with the CoT-SQL outputs from a teacher model (a TTS SQL Gen LLM) to form a distillation dataset of NL(-\ukf)-CoT-SQL tuples. This supports the student model's lifelong learning through either SFT on the tuples, or direct SFT by ignoring the CoT in the accumulated data.

The quality and selection policy for this accumulated data are critical. In Section.~\ref{sec:method-inf-tts}, we explore TTS techniques to improve the teacher model's generation quality. We also introduce \emph{Cascading}, a related TTS method designed to build an efficient workflow with the student model(s). Section.~\ref{sec:method-inf-data} details our process for creating a high-quality, selected subset of NL-CoT-SQL tuples from the accumulated data.

\subsubsection{SQL Generation with TTS and Cascading}
\label{sec:method-inf-tts}

Test-Time Scaling (TTS)~\cite{TTS} refers to a set of LLM inference policies that improve performance without fine-tuning, primarily by increasing the total number of tokens generated during inference(s). The use of CoT, which prompts the LLM to produce intermediate reasoning tokens before the final answer, is itself considered a common TTS technique.

Another standard TTS approach involves ensembling, verifying, and selecting from multiple parallel LLM outputs. \nlsql is particularly well-suited for such techniques because its output—SQL—can be easily verified for \emph{executability}. Therefore, \sys employs \emph{majority voting with SQL execution verification} as its primary TTS scheme. Specifically, it first samples $n > 1$ SQL outputs, discards any that fail to compile, and then performs a majority vote on the execution results of the remaining queries. A tie-breaker can be configured by user preference or by using an LLM to analyze the SQL queries directly. Notably, this TTS method is model-agnostic, enabling the ensembling of heterogeneous models with varying hyperparameters, sizes, or architectures.

While TTS improves the teacher model's quality, we use \emph{Cascading} to address latency of the student model. Like TTS, cascading also leverages multiple parallel LLM inferences, but its goal is to reduce overall latency by applying TTS to smaller, faster student models to replace a slower model. This is because in enterprise \nlsql scenarios, user queries are often sparse over time, resulting in low throughput requirements, yet low latency remains critical for user experience. With idle computation resources and a strict latency requirement, \sys can trade idle throughput for accurate SQL generation with lower latency. Specifically, it executes multiple LLMs of different sizes (e.g., 14B, 32B, 671B) in parallel, with each model generating at least $n = 2$ samples. As soon as two matching execution results are detected, the cascading system returns the result and terminates all remaining threads. For instance, if the two 14B model outputs agree, that result is returned immediately, bypassing the need to run the larger 32B and 671B models. If the 14B outputs disagree, the 32B model acts as a judge, and its output is returned if it agrees with one of the 14B proposals, and so on.

Another direction worth investigating is \emph{continued inference}. As RLMs generate CoT-SQL pairs, we observe that complex, commented SQL is on average $2\times$ longer than the CoT that precedes it. However, generating SQL is relatively straightforward once the logical reasoning and sketch are provided by the CoT. Therefore, latency can be further reduced by using a relatively large model to generate the CoT and a relatively small model to complete the SQL generation using the large model's outputs as its own output prefix (an approach sometimes called ``assistant prefill''). We leave the exploration of continued inference and its integration with cascading — treating it as another model type in the ensemble — as future work.

\subsubsection{Knowledge Distillation}
\label{sec:method-inf-data}

Knowledge distillation~\cite{cotd} (under LLM fine-tuning context) means constructing training datasets with CoTs to transfer the reasoning ability of LLMs to small LMs. This approach mainly consists of two stages: training data curation, and the actual post-training. As demonstrated in previous knowledge distillation works~\cite{cotd,s1K}, the quality of reasoning traces is more crucial than their quantity. Therefore, selecting correct, high-quality data for training is vital. This section proposes a scoring mechanism specifically designed to evaluate NL-CoT-SQL tuples. Similar to s1K~\cite{s1K}, our scoring system has three categories: hardness score $\mathcal{H}$, quality score $\mathcal{Q}$, and variety score $\mathcal{V}$. For each NL-CoT-SQL tuple, the total score is:
\begin{equation}
    \mathcal{S} = \alpha\mathcal{H} + \beta\mathcal{Q} + \gamma\mathcal{V} + \mathcal B
\end{equation}
where $\alpha, \beta, \gamma \in [0,1]$ are coefficients for balancing the scores, and $\mathcal B$ is an arbitrary tie-breaker function.

\paragraph{Hardness Score $\mathcal{H} \in [0,N]$.} During training data curation, we evaluate both advanced LLMs (e.g., DeepSeek-671B) and small LMs (e.g., Qwen-32B) on the same training data. If $N$ different models or sampled trajectories are included, the hardness of a data point is defined as the number of LLMs that failed to produce a correct SQL query. This approach penalizes simple queries, leading to the selection of more challenging queries for the training dataset.

\paragraph{Quality Score $\mathcal{Q} \in [0,70]$.} This score includes seven dimensions, each contributing $0 \sim 10$ points to reward CoT quality. These dimensions are: information completeness, robustness (corner cases), SQL structure clarity, example referencing, structured thinking, non-repetitiveness, and brevity. The non-repetitiveness and brevity scores are included for latency considerations, penalizing long reasoning traces and encouraging succinct problem-solving. The brevity score is obtained simply by counting the number of tokens, while other scores are generated by an LLM judge, which is explicitly instructed to generate results as close as possible to a \emph{normal distribution}.

\paragraph{Diversity Score $\mathcal{V}$.} The diversity score can be defined either in domain-specific or domain-agnostic fashion. With proper domain knowledge, it encourages queries with lower frequencies to mitigate undersampling. Adjusting the diversity score can be an ad-hoc task. Alternatively, the score can be generated by clustering training queries and scoring their diversity based on their average distance to their $k$-closest neighbors.

Finally, the NL-CoT-SQL tuples are sorted according to the total score $\mathcal{S}$. Only the top-scored tuples are then selected for the knowledge distillation training set.

After obtaining the curated training set, for post-training, we directly perform full-parameter SFT to train small LMs (e.g., 14B, 32B) based on the instruct version (e.g., \texttt{Qwen2.5-14B/32B-Instruct}) of the models~\cite{s1K}. As described in Section.~\ref{sec:method-inf-tts}, it is also possible to consider knowledge distilled models as parts of a heterogeneous ensemble for TTS.

\end{sloppypar}

%% file: src/benchmark.tex
\section{\bench Benchmark}
\label{sec:dataset}
\begin{sloppypar}

The development of \nlsql benchmarks such as Spider~\cite{spider}, BIRD~\cite{BIRD}, and WikiSQL~\cite{WikiSQL} has been indispensable for advancements in \nlsql technology. While these datasets have been designed to evaluate the boundaries of general-purpose \nlsql tasks, including multi-hop reasoning and table discovery capabilities, their direct applicability to real-world enterprise scenarios is limited; for such scenarios, domain-specific challenges like robust schema linking and the interpretation of implicit user intents prove to be more crucial problems. Most importantly, these benchmarks, as well as recent agent-oriented benchmarks~\cite{spider2,birdinteract,livesql}, are more of wide-ranging rather than specialized, often stuffed with many different databases but few query types per database.

Benchmarks like BIS~\cite{BIS} and BEAVER~\cite{beaver} represent valuable steps toward addressing business applications by focusing primarily on enterprise scenario core databases with complex wide-table schemas. However, these benchmarks suffer from an insufficient number of questions, which constrains the diversity of background knowledge and impedes their ability to comprehensively evaluate lifelong learning systems.

As mentioned in Section~\ref{sec:intro}, we identified key challenges in real-world NL2SQL applications that existing benchmarks rarely address, i.e. implicit intent, private domain knowledge, wide table schemas, and context sensitivity. To address these gaps, we introduce \bench, the first large-scale benchmark specifically focused on enterprise finance \nlsql scenarios. \bench is meticulously derived from real-world business queries, thereby ensuring realistic question distributions over complex schemas and domain-specific challenges. The source queries are extracted from a production \nlsql system currently in use within the finance departments of a multinational enterprise, with anonymized queries posed by finance analysts, administrative staff, and C-suite executives (CFOs, CEOs). This diverse origin captures a wide spectrum of enterprise Business Intelligence (BI) demands.

\subsection{Benchmark Characteristics}
\label{sec:dataset-traits}
The \bench database comprises four table categories related to a hypothetical car manufacturer: ``Income'', ``Sales Ledger'', ``Profit and Loss'', and ``Budget and Forecast''. The automobile company contains 8 departments; each department has its own ``Income'' and  ``Sales Ledger'' table, while all sharing ``Profit and Loss'' and ``Budget and Forecast''. Each table has $\sim 40K$ records and $\sim 70$ columns. To ensure confidentiality and data privacy, we employ robust privacy-preserving approaches, desensitizing sensitive information while rigorously retaining structural fidelity.

\benchI currently contains over 5,000 annotated queries compatible with PostgreSQL and SQLite. The query set will undergo iterative refinement, aligning with RUBIKSQL's core philosophy: SQL generation and KB maintenance should form an interleaving process to continuously enhance system performance. Queries in \bench are constructed and improved through a four-stage approach:
\begin{itemize}
    \item \textbf{Synthetic Seed Query Generation}: An initial knowledge base $\mathcal{K}^0$ is established by generating synthetic seed queries from the database schema.
    \item \textbf{Real-World Query Transfer}: Real-world business queries from a private dataset are transferred to the database, similar to the query transfer method discussed in Section.~\ref{sec:method-aug-query}.
    \item \textbf{Query Augmentation}: Additional queries are generated through rephrasing and composition, then answered by powerful LLMs with the aid of the updated knowledge base, mirroring the approach in Section.~\ref{sec:method-mining-agent}.
    \item \textbf{Manual Verification}: All generated queries undergo rigorous manual examination by financial experts and skilled SQL data analysts to ensure correctness and diversity, forming the final benchmark.
\end{itemize}

\bench places strong emphasis on \emph{context-aware queries}. Each query in \bench is associated with a specific organizational role, ranging from finance analysts who use precise metrics and carefully formulated questions, to executives who pose high-level, often ambiguous requests that necessitate intent inference. This makes \bench particularly suited for evaluating the NL2SQL utility in real-world enterprise settings. For instance, the benchmark incorporates vague requests like ``Revenue by region?'', which lacks explicit region definitions, mirroring complexities typical in actual enterprise environments. 
To resolve such ambiguities, in addition to database content, schema, and queries, \bench provides supplementary contextual resources, including user profiles, documents of business jargons and descriptions of internal rules.

One defining characteristic of \bench lies in its source of queries: transferred queries derived from real-world deployed workloads. This fundamentally distinguishes it from conventional synthetic or expert-labeled benchmarks. Direct release of real enterprise datasets -- particularly in sensitive domains such as financial and management systems, is exceedingly rare due to serious security concerns regarding information leakage. Furthermore, developing advanced data synthesis approaches that accurately capture the intricacies of real enterprise environments remains a major challenge, often demanding substantial domain expertise and effort. By leveraging realistically transferred queries, \bench fills this critical gap and provides a robust evaluation resource for \nlsql systems targeting enterprise scenarios. It effectively mitigates the limitations of purely synthetic datasets and the scarcity of authentic publicly available enterprise data in NL2SQL world.

\subsection{Performance Metrics}
\label{sec:dataset-metrics}

As noted by BIRD Mini-Dev and LiveSQLBench~\cite{BIRD,livesql}, the exact Execution Accuracy (EX) score can be unnecessarily strict for evaluating \nlsql systems. In practice, an SQL query may produce nearly-correct results -- containing all required information, yet present it in a format differing from the ground-truth. Examples include variations in column order or the presence 
of additional columns
(e.g., a ``rank'' column in a sorted result). To mitigate this, BIRD Mini-Dev introduced a \textit{soft $F_1$-score}, which averages the $F_1$-scores computed between corresponding rows treated as \emph{unordered dictionaries}.

We also observed that in tasks such as aggregation, variations in column names can significantly impact evaluation; for example, comparing \texttt{SELECT revenue - cost} with \texttt{SELECT revenue - cost AS `profit'}. Furthermore, execution result row order matters only when the ground-truth SQL contains an \texttt{ORDER BY} clause. The issue of extraneous columns is already handled by the soft $F_1$-score, though the presence of extra rows can also affect evaluation, especially with omitted rows or summary rows. Last but not least, in practical \nlsql applications, recall is often considered more important than precision, as the primary user objective is to retrieve all relevant information. This motivates the definition of a \textit{Bipartite $F_\beta$-score} ($\text{BF}_{\beta}$) as a refined metric for evaluating SQL execution, with $\beta > 1$. 

Technically, we compare the predicted SQL's execution result, which is an ordered list of columns value sets (column names removed) $P = \langle p_1, p_2, \dots, p_n \rangle$, with the ground-truth SQL's execution result $G = \langle g_1, g_2, \dots, g_m \rangle$ by defining the \textit{row-wise $F_\beta$-score} as:
\begin{align}
w^\beta_{i,j} = \frac{(1+\beta^2) \cdot \text{Pre}_{i,j} \cdot \text{Rec}_{i,j}}{\beta^2 \cdot \text{Pre}_{i,j} + \text{Rec}_{i,j}} \\
\text{Pre}_{i,j} = \frac{1}{|p_i|}\sum_{c \in p_i} \mathbb{I}[c \in g_j] \\
\text{Rec}_{i,j} = \frac{1}{|g_j|}\sum_{c \in g_j} \mathbb{I}[c \in p_i]
\end{align}
where $\mathbb{I}$ is the indicator function. This formulation creates a fully-connected weighted bipartite graph $\boldsymbol{w}^\beta(P,G)$ indicating row-pair similarities.

Next, we calculate the best match between the prediction and ground-truth using this weighted bipartite graph. For unordered queries (where the ground-truth SQL has no \texttt{ORDER BY} clause), the score is calculated using a Weighted Bipartite Matching ($WBM$) via the Hungarian algorithm~\cite{hungarian}. For ordered queries (where the ground-truth SQL includes an \texttt{ORDER BY} clause), the calculation of the score is a non-intersecting maximum weighted bipartite matching ($WBM_{NI}$) task, which can be computed via dynamic programming — with $DP(i,j)$ being the maximum matching considering the first $i$ rows in $P$ and first $j$ rows in $G$. If we denote $N = \max(\vert P \vert, \vert G \vert)$, the time complexity of the Hungarian Algorithm for $WBM$ is upper-bounded by $O(N^3)$, but in most cases, the ``fully-connected'' weights are actually sparse with many $w=0$ and almost 1-to-1 match, so the average runtime is roughly linear. Meanwhile, the dynamic programming has a bounded complexity of $O(\vert P \vert \cdot \vert G \vert) \le O(N^2)$, and roughly linear using recursive DP with greedy pruning.

Finally, the \textit{Bipartite $F_\beta$-score} ($\text{BF}_{\beta}$) is defined as:
\begin{equation}
\begin{split}
\text{BF}_{\beta}(P,G) = \frac{1}{|Q|}\left(\sum_{P,G \in \text{Eval}(Q_\text{unordered})} \frac{WBM(\boldsymbol{w}^\beta(P,G))}{\max(|P|, |G|)} \right. \\
\left. + \sum_{P,G \in \text{Eval}(Q_\text{ordered})} \frac{WBM_{NI}(\boldsymbol{w}^\beta(P,G))}{\max(|P|, |G|)} \right)
\end{split}
\end{equation}

The \textit{Bipartite $F_\beta$-score} offers several key benefits. It accurately handles differences in column order, extra columns, and extra rows (including omitted rows and summary rows), which are common issues in \nlsql tasks. Furthermore, because it uses the $F_\beta$-score, the $\beta$ value can be adjusted, with $\beta > 1$ to emphasize recall over precision, aligning with practical user needs. Lastly, this metric supports evaluation across diverse result sets, making it a robust and flexible tool for assessing \nlsql system performance.
\end{sloppypar}

%% file: src/exp.tex
\section{Experiment}
\label{sec:exp}

To demonstrate the effectiveness of \sys, we performed experiments on open-source \nlsql benchmarks. Considering \sys's transductive and lifelong learning nature, we focused our evaluation on benchmarks with queries concentrated on a limited number of databases. This is a more suitable approach than using datasets with sparse queries spread across many databases, which are typically used for assessing generalizability. Many existing \nlsql benchmarks, such as WikiSQL~\cite{WikiSQL} and Spider~\cite{spider}, feature a wide variety of databases with relatively few queries per database, making them less suitable for evaluating lifelong learning. Others, like BookSQL~\cite{BookSQL} and SEDE~\cite{SEDE}, have not been actively maintained. This motivated our development of the new \bench benchmark. Nevertheless, to provide a broader comparison and validate \sys's performance on existing datasets, we also conducted evaluations on KaggleDBQA~\cite{KaggleDBQA} and BIRD Mini-Dev~\cite{BIRD}. KaggleDBQA contains 8 distinct databases, while BIRD Mini-Dev includes 19 databases. Both offer a denser query distribution compared to general-purpose benchmarks like WikiSQL.

\subsection{Baselines}
\label{sec:exp-baselines}

On BIRD, we compare with publicly reported Dev Execution Accuracy (EX) figures of leading methods such as CSC-SQL~\cite{CSC}, XiYan-SQL~\cite{XiYanSQL}, CYAN-SQL, Contextual-SQL~\cite{contextual}, TCDataAgent-SQL, CHASE-SQL~\cite{CHASE}, AskData~\cite{AskData} and LongData-SQL. On KaggleDBQA, we compare with publicly reported Test EX figures, including RAT-SQL~\cite{RAT-SQL}, DIN-SQL~\cite{DIN-SQL}, ZeroNL2SQL~\cite{ZeroNL2SQL}, and ODIS-Codex~\cite{ODIS}. All methods are LM-based, using either finetuned LMs (32B) or LLMs/agents with TTS.

\subsection{Settings}
\label{sec:exp-settings}

As the open-source benchmarks do not possess some traits of an enterprise data (e.g., user identity, time-sensitivity, etc.), we perform a selective subset of the aforementioned \sys techinques according to the benchmark charateristics. Specifically, on both BIRD Mini-Dev and KaggleDBQA, since no documentation was available, we employed database profiling and agentic context mining for database context engineering on each database. Only the \textit{Enum, Predicate, Column, Table}, and \textit{Special} instances of \ukf were used during knowledge retrieval. The KB index included the LLM-augmented DAAC index and the multi-vector index, in which we used \emph{query sketch} method as the serializer for \textit{Experience} and \texttt{name} as a naive serializer for other \ukf instances. For BIRD Mini-Dev, we applied \emph{query transfer} to leverage the large number of queries from training databases; for KaggleDBQA, we did not apply query transfer since the training and test sets share the same databases. We did not apply knowledge distillation for these two benchmarks, and applied only the simplest SQL profiling (query information and AI-fiendly SQL formatting). We used \texttt{all-MiniLM-L12-v2} (\texttt{all-minilm:33m}) as our embedding model. We used \texttt{gemini-2.5-flash} (\texttt{preview-05-20}) as the LLM for all agents and for all LLM-based operations. For small LMs, we used \texttt{DeepSeek-R1-Distill-Qwen-32B} and \texttt{Qwen2.5-72B-Instruct}. For our TTS choices, we conducted our experiments with both $n=1$ and \emph{majority voting with SQL execution verification} with $n=8$.

\subsection{Main Results}
\label{sec:exp-results}

\begin{table}[htbp]
    \centering
    \caption{Results on the BIRD Dev/Mini-Dev Dataset}
    \label{tab:bird_dev_results}
    \begin{tabularx}{0.81\linewidth}{l c c}
        \toprule
        \textbf{Method} & \textbf{TTS} & \textbf{Dev EX (\%)} \\
        \midrule
        \texttt{DeepSeek-R1} & n=1 & 52.0\footnotemark[1] / 56.13 \\
        \texttt{gemini-2.5-flash} & n=1 & 59.4\footnotemark[1] \\
        \midrule
        CSC-SQL & n=72 & 71.33 \\
        XiYan-SQL & n=5 & 73.34 \\
        CYAN-SQL & unknown & 73.47 \\
        Contextual-SQL & n=32 & 73.50 \\
        TCDataAgent-SQL & unknown & 74.12 \\
        LongData-SQL & unknown & 74.32 \\
        CHASE-SQL & n=21 & 74.90 \\
        AskData & n=3 & 73.0\footnotemark[1] / 75.36 \\
        \midrule
        \textbf{\sys (Ours)} & n=1 & 75.9\footnotemark[1] \\
        \textbf{\sys (Ours)} & n=8 & \textbf{77.3}\footnotemark[1] \\
        \bottomrule
    \end{tabularx}
\end{table}
\footnotetext[1]{Evaluated on BIRD Mini-Dev, which is a subset of BIRD Dev with less queries. Other figures are evaluted on BIRD Dev.}

\begin{table}[htbp]
    \centering
    \caption{Results on the KaggleDBQA Dataset}
    \label{tab:kaggledbqa_results}
    \begin{tabularx}{0.63\linewidth}{l c c}
        \toprule
        \textbf{Method} & \textbf{TTS} & \textbf{EX (\%)} \\
        \midrule
        RAT-SQL & n=1 & 26.8 \\
        DIN-SQL & n=1 & 27.0 \\
        ZeroNL2SQL & n=1 & 44.9 \\
        ODIS-Codex & n=1 & 54.8 \\
        \midrule
        \textbf{\sys (Ours)} & n=1 & 54.1 \\
        \textbf{\sys (Ours)} & n=8 & \textbf{58.9} \\
        \bottomrule
    \end{tabularx}
\end{table}

On both BIRD Mini-Dev and KaggleDBQA, \sys achieves SOTA performance with $n=8$ TTS, demonstrating the advantage of \sys system. To further examine the effectiveness of \sys, we tested on BIRD Mini-Dev with \texttt{gemini-2.5-flash} (Table.~\ref{tab:bird_dev_results}, the same model as used in our experiemnt, using only DDL), which yields a performance of $59.4\%$ without \sys.

\begin{figure}
    \centering
    \includegraphics[width=0.80\linewidth]{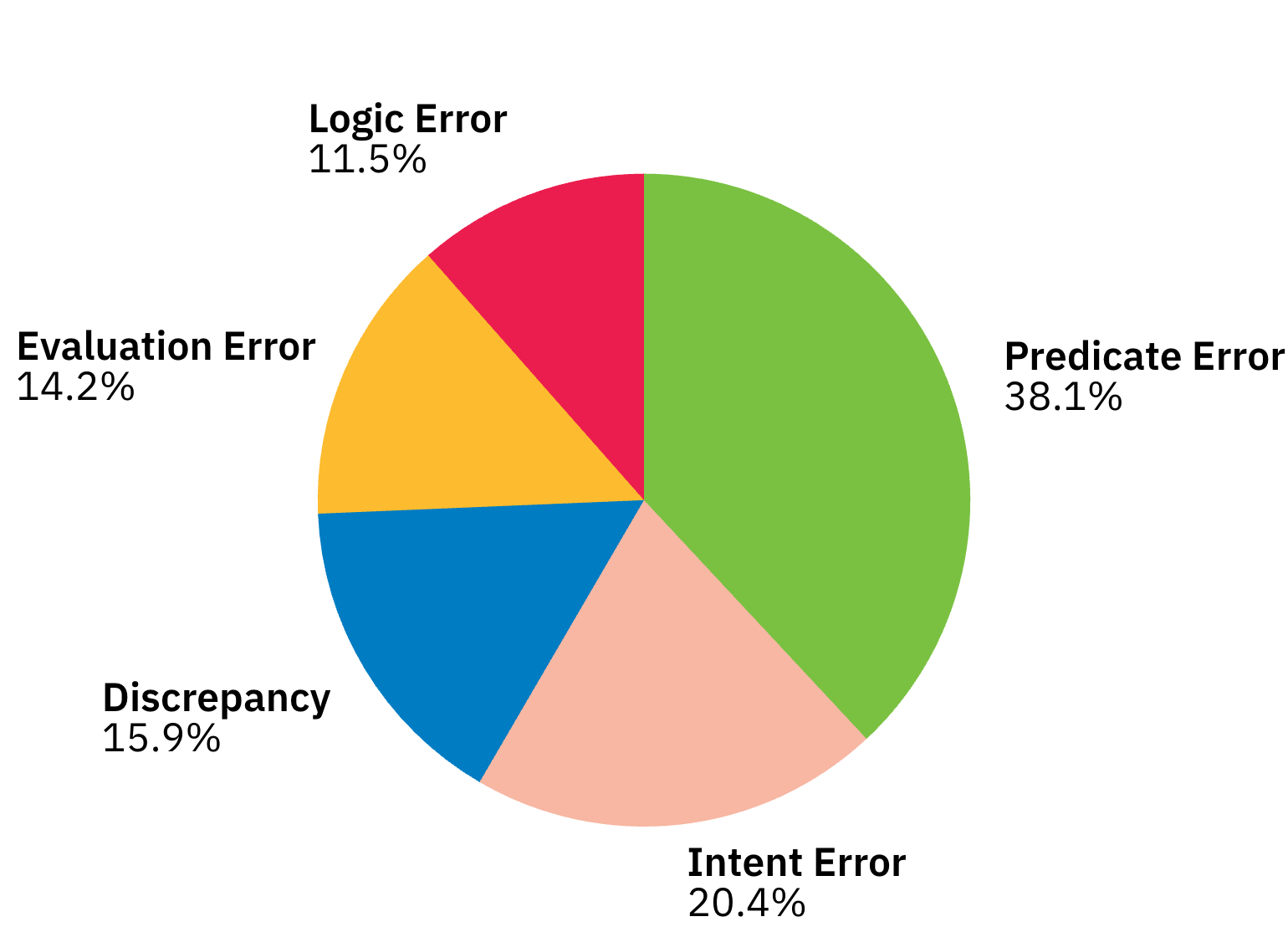}
    \caption{Error Type Analysis of \sys on BIRD Mini-Dev.}
    \Description{Error Type Analysis of \sys on BIRD Mini-Dev.}
    \label{fig:error}
\end{figure}

A more detailed error analysis is carried out on BIRD Mini-Dev and presented in Table.~\ref{fig:error} and Figure.~\ref{fig:error}. BIRD classifies its queries into three difficulty classes: ``simple'', ``moderate'', and ``challenging'', with a rough distribution of $30\%$-$50\%$-$20\%$:

\begin{table}[h]
\centering
\caption{Error Distribution by Difficulty Class}
\label{tab:error-analysis}
    \begin{tabularx}{0.99\linewidth}{lcccc}
        \toprule
                        & \multicolumn{3}{c}{\textbf{Difficulty}} & \\
        \cmidrule(lr){2-4}
        \textbf{Metric} & \textbf{Simple} & \textbf{Moderate} & \textbf{Challenging} & \textbf{Overall} \\
        \midrule
        Ratio (\%) & 29.8 & 49.7 & 20.5 & 100.0 \\
        EXAcc (\%) & 85.1 & 76.1 & 68.6 &  77.3 \\
        \midrule
        Error Dist (\%) & 19.5 & 52.2 & 28.3 & 100.0 \\
        Dist Diff (\%) & -10.3 & +2.5 & +7.8 & -- \\
    \bottomrule
    \end{tabularx}
\end{table}

The error distribution of \sys on BIRD closely mirrors the original query distribution, with only a $+7.8\%$ delta in the error rate for the challenging class. This indicates that \sys's performance scales with query complexity and does not exhibit a specific weakness in handling challenging queries or any other particular difficulty class. Given that we use only the \texttt{gemini-2.5-flash} model, this result supports our claim that a primary bottleneck in \nlsql accuracy is \emph{NOT} the LLM or agent capability, but rather the quality of the underlying KB.

Figure.~\ref{fig:error} shows the distribution of error types. Predicate errors, which constitute $\sim 40\%$ of the total errors, include SQLs with mostly correct computational logic but with extra or missing predicates ($15.9\%$), or incorrect enumerations ($7.1\%$) or column names ($7.9\%$) in the predicates. Predicate errors also involve misinterpreting data formats ($7.1\%$), particularly \texttt{datetime} values ($5.3\%$). For example, range query over to a datetime column stored as non-ISO strings. The prevalence of predicate errors suggests a significant opportunity to further improve KB utilization for better information presentation to the agents, including mechanisms to enforce checks for correct enumerations and data formats.

Approximately 20\% of errors stem from a failure to accurately interpret user intent. This includes cases where the generated SQL retrieves extra columns or less columns than expected (7.1\%) and improper post-processing, such as a missing conversion from ratios to percentages or over-processing like string field concatenation (4.4\%). Another significant source of error is disagreement over tie-breaking for \texttt{ORDER BY} or \texttt{LIMIT} clauses (8.8\%). These intent-related errors can \emph{NOT} be solved solely by improving the base model's capabilities. Instead, they require either aligning the \nlsql system with the majority's preference or, more effectively, constructing user profiles from extra knowledge input or historical queries to address individual user preferences, as we have advocated.

Discrepancies in data content account for $\sim 16\%$ of errors. These include disagreements over \texttt{NULL} handling, \texttt{DISTINCT}, multiple valid ways of computation (e.g., using `\texttt{AVG}` vs. `\texttt{SUM}/\texttt{COUNT}`), or the same value being stored in different columns with minor discrepancies. Such errors can stem from insufficient prompting, LLM capabilities, or poor database consistency. While they do not significantly affect user experience during large-scale statistical analysis, they can cause issues with local, narrow queries.

Approximately $14\%$ of errors lie in the benchmark itself. These include erroneous ground truth SQLs, potential floating-point inaccuracies during evaluation, and disagreements over the undefined ordering of rows and columns.

Only less than $12\%$ of errors are caused by insufficient LLM reasoning, resulting in missing computational logic or completely incorrect solutions. This low percentage once again supports our primary claim that the main bottleneck in \nlsql accuracy lies in the quality of the underlying KB rather than LM capabilities.

In conclusion, the comprehensive error analysis reveals that the primary bottlenecks in the \sys's performance are not due to the LLM or agent's reasoning capabilities. Instead, the majority of errors stem from issues related to the quality of the resentation of information to the agents, and understanding of user intents, suggesting the need of incorporating user profiles and historical query data. This analysis validates our core claim that for industrial NL2SQL systems, the KB is the most critical component.

%% file: src/limit.tex
\section{Limitations and Future Works}
\label{sec:limit}
\begin{sloppypar}

\sys presents a systematic, end-to-end solution for enterprise-level \nlsql, with a focus on practical deployment. While our work is among the first to offer such an integrated and systematic approach, introducing novel insights and optimizations for specific modules — including the \ukf semantic layer, the LLM-augmented DAAC index, CoT-enhanced SQL profiling, and training data curation, we must acknowledge that many components merit further investigation. We leave several modules with established solutions (e.g., structured information extraction) and others with simplified implementations (e.g., majority vote, round-robin, graph-based index, and post-training) or contributions that require more detailed ablation studies (e.g., query synthesis and transfer). We plan to add detailed ablation experiemnts in the near future.

Meanwhile, though our multi-agent workflow effectively demonstrates the power of the KB, its primary purpose is to serve as a built-in example rather than a fully-optimized solution — as the core concept of \sys is to \emph{reduce} the reliance on agentic abilities by building increasingly comprehensive and AI-friendly KBs.

Finally, as the \benchI is still in its primitive stages, we consider its optimization, experiments, along with the improvement of individual \sys components and a detailed survey of modules, as promising directions for future work. Our work highlights the critical importance of knowledge and systematic lifelong learning in \nlsql and provides a robust foundation for the community to build upon.

\end{sloppypar}

%% file: src/conclusion.tex
\section{Conclusion}
\label{sec:conclusion}
\begin{sloppypar}

In this paper, we introduce \sys, a novel \nlsql system designed to handle the complexities of real-world enterprise environments, including implicit intent, private domain knowledge, wide table schema and context sensitivity. The \sys system contains a four-stage knowledge-centric agentic workflow: \emph{Database Context Engineering}, \emph{User Query Augmentation}, \emph{Knowledge Base Indexing}, and \emph{Knowledge Distillation}. These stages are bridged by the \emph{Unified Knowledge Format} (\ukf), a crucial semantic layer that decouples knowledge extraction and maintainance from its utilization. Together, these components from as an \emph{Agentic Knowledge Base}, functioning as a whole as the core of a lifelong learning \nlsql system. Our SOTA experimental results on the KaggleDBQA~\cite{KaggleDBQA} and BIRD Mini-Dev~\cite{BIRD} datasets demonstrate the superiority of the \sys system. We are also releasing \bench, a new benchmark tailored to the unique characteristics of industrial \nlsql, including a single database with realistic schemas and \emph{context-aware queries}, to foster further research in this domain and call for more research attention to knowledge-centric \nlsql systems.

\end{sloppypar}

%% file: src/appendix.tex
\appendix

\section{Unified Knolwedge Format}
\label{sec:appendix-ukf}
\begin{sloppypar}

\subsection{\ukfI Definition}
\label{sec:appendix-ukf-def}

UKF divides knowledge attributes into six core groups, each containing multiple fields, together forming a comprehensive and flexible knowledge representation model:

\subsubsection{Metadata}

This section contains descriptive attributes, providing basic information about the knowledge, including human-readable and LLM-interpretable descriptions.

\begin{itemize}
    \item \textbf{\texttt{name}}: (Required) Unique identifier for the knowledge. This field is immutable after creation.
    \begin{itemize}
        \item Type: string
    \end{itemize}

    \item \textbf{\texttt{notes}}: (Optional) Human-readable description of the knowledge, used for detailed explanations or context.
    \begin{itemize}
        \item Type: string
        \item Default: empty string \texttt{""}
    \end{itemize}
    
    \item \textbf{\texttt{short\_description}}: (Optional) A brief description for LLMs, intended to provide a concise summary.
    \begin{itemize}
        \item Type: string
        \item Default: empty string \texttt{""}
    \end{itemize}
    
    \item \textbf{\texttt{description}}: (Optional) A detailed description for LLMs, providing more comprehensive information.
    \begin{itemize}
        \item Type: string
        \item Default: empty string \texttt{""}
    \end{itemize}
    
    \item \textbf{\texttt{type}}: (Optional) Knowledge type identifier, used to distinguish different categories of knowledge. Immutable after creation.
    \begin{itemize}
        \item Type: string
        \item Default: \texttt{"general"}
        \item Example: \texttt{"finance"}, \texttt{"healthcare"}, \texttt{"education"}, ...
    \end{itemize}
    
    \item \textbf{\texttt{version}}: (Optional) Version number of the knowledge schema, usually auto-filled. Immutable after creation.
    \begin{itemize}
        \item Type: string
        \item Default: \texttt{"v0.1.0"}, \texttt{"v0.2.2a"}
    \end{itemize}
    
    \item \textbf{\texttt{version\_notes}}: (Optional) Human-readable description of version changes.
    \begin{itemize}
        \item Type: string
        \item Default: empty string \texttt{""}
    \end{itemize}
    
    \item \textbf{\texttt{variant}}: (Optional) Variant identifier, e.g., different languages, models, or contexts. Immutable after creation.
    \begin{itemize}
        \item Type: string
        \item Default: \texttt{"default"}
    \end{itemize}
    
    \item \textbf{\texttt{variant\_notes}}: (Optional) Human-readable description of the variant.
    \begin{itemize}
        \item Type: string
        \item Default: empty string \texttt{""}
    \end{itemize}
\end{itemize}

\subsubsection{Content}

This section defines the core content of the knowledge, supporting text and semi-structured representations.

\begin{itemize}
    \item \textbf{\texttt{content}}: (Optional) Main knowledge payload, usually in text form. Immutable after creation.
    \begin{itemize}
        \item Type: string
        \item Default: empty string \texttt{""}
    \end{itemize}
    
    \item \textbf{\texttt{content\_resources}}: (Optional) Auxiliary assets or context to supplement the main content. Immutable after creation.
    \begin{itemize}
        \item Type: Dict[str, Any]
        \item Default: empty dict \texttt{\{\}}
    \end{itemize}
    
    \item \textbf{\texttt{content\_composers}}: (Optional) Content processing functions for dynamic serialization, supporting context-, language-, and model-aware prompt construction. Immutable after creation.
    \begin{itemize}
        \item Type: Dict[str, Callable], keys are strings, values are callables.
        \item Default: includes a default composer function named \texttt{"default"}, each composer takes the knowledge and arbitrary kwargs as arguments and returns a string: \texttt{f(kl:BaseUKF, **kwargs) -> str}.
        \item Example: \texttt{\{"default": (lambda kl, **kwargs: kl.content), "custom": (lambda kl, **kwargs: kl.content.upper())\}}
    \end{itemize}
\end{itemize}

\subsubsection{Provenance}

This section contains source tracking attributes, supporting personalized knowledge bases and privacy/permission control.

\begin{itemize}
    \item \textbf{\texttt{source}}: (Optional) Source of the knowledge. Immutable after creation.
    \begin{itemize}
        \item Type: \texttt{Literal['system', 'user', 'auto', 'tool', 'derived', 'unknown']}
        \item Default: \texttt{'unknown'}
    \end{itemize}
    
    \item \textbf{\texttt{parents}}: (Optional) Source knowledge identifiers for tracing origin or dependencies. Immutable after creation.
    \begin{itemize}
        \item Type: Dict[str, Any]
        \item Default: empty dict \texttt{\{\}}
    \end{itemize}
    
    \item \textbf{\texttt{owner}}: (Optional) Identifier of the knowledge owner (usually a user). Immutable after creation.
    \begin{itemize}
        \item Type: string
        \item Default: \texttt{'unknown'}
    \end{itemize}
    
    \item \textbf{\texttt{workspace}}: (Optional) Context namespace or scope. Immutable after creation.
    \begin{itemize}
        \item Type: string
        \item Default: \texttt{'unknown'}
    \end{itemize}
    
    \item \textbf{\texttt{creator}}: (Optional) Identifier of the creator (user/agent/tool). Immutable after creation.
    \begin{itemize}
        \item Type: string
        \item Default: \texttt{'unknown'}
    \end{itemize}
\end{itemize}

\subsubsection{Retrieval}

This section contains attributes for optimizing search indexing and recommendation systems.

\begin{itemize}
    \item \textbf{\texttt{collection}}: (Optional) Identifier for the knowledge collection, used for grouping. Immutable after creation.
    \begin{itemize}
        \item Type: string
        \item Default: \texttt{"general"}
    \end{itemize}
    
    \item \textbf{\texttt{tags}}: (Optional) List of tags for classification, unordered. Immutable after creation.
    \begin{itemize}
        \item Type: Set[str]
        \item Default: empty set \texttt{set()}
        \item Example: \texttt{["[TOPIC:Artificial Intelligence]", "[USER\_GROUP:Researchers]"]}
    \end{itemize}
    
    \item \textbf{\texttt{synonyms}}: (Optional) Alternative names for keyword matching.
    \begin{itemize}
        \item Type: Set[str]
        \item Default: empty set \texttt{set()}
    \end{itemize}
    
    \item \textbf{\texttt{triggers}}: (Optional) Condition triggers for activating the knowledge.
    \begin{itemize}
        \item Type: Dict[str, Callable], keys are strings, values are callables.
        \item Default: includes a default trigger function named \texttt{"default"}, each trigger takes the knowledge and arbitrary kwargs as arguments and returns a boolean: \texttt{f(kl:BaseUKF, **kwargs) -> bool}.
        \item Example: \texttt{\{"default": (lambda kl, **kwargs: "sports" in kwargs.get("question",""))\}}
    \end{itemize}
    
    \item \textbf{\texttt{priority}}: (Optional) Priority of the knowledge, only as an attribute, not directly used by the knowledge itself, the knowledge base, or search engines. Immutable after creation.
    \begin{itemize}
        \item Type: int
        \item Default: \texttt{0}
    \end{itemize}
\end{itemize}

\subsubsection{Relationships}

This section defines knowledge graph tuples, where predicates can refer to other UKF instances.

\begin{itemize}
    \item \textbf{\texttt{related}}: (Optional) Relationships with other knowledge. Each relationship is a tuple \texttt{(subject\_id: int, relation: str, object\_id: int, relation\_id: Optional[int], relation\_resources: Optional[Dict[str, Any]])}, representing subject, predicate, object, relation knowledge ID (optional), and relation resources (optional).
    \begin{itemize}
        \item Type: Set[Tuple[int, str, int, Optional[int], Optional[Dict[str, Any]]]]
        \item Default: empty set \texttt{set()}
        \item Example: \texttt{\{(<id1>, "is\_a", <id2>, None, None)\}}
    \end{itemize}
    
    \item \textbf{\texttt{auths}}: (Optional) User permissions, each as a tuple \texttt{(user, authority)}.
    \begin{itemize}
        \item Type: Set[Tuple[str, str]]
        \item Default: empty set \texttt{set()}
    \end{itemize}
\end{itemize}

\subsubsection{Life-cycle}

This section contains time management attributes, supporting dynamic updates and version control.

\begin{itemize}
    \item \textbf{\texttt{timefluid}}: (Optional) Marks whether the knowledge is time-sensitive. Immutable after creation.
    \begin{itemize}
        \item Type: bool
        \item Default: \texttt{False}
    \end{itemize}
    
    \item \textbf{\texttt{timestamp}}: (Optional) Auto-generated creation timestamp.
    \begin{itemize}
        \item Type: datetime.datetime
        \item Default: current time
    \end{itemize}
    
    \item \textbf{\texttt{last\_verified}}: (Optional) Last verification time.
    \begin{itemize}
        \item Type: datetime.datetime
        \item Default: current time
    \end{itemize}
    
    \item \textbf{\texttt{expiration}}: (Optional) Expiration time (seconds since last update). \texttt{-1} means never expires.
    \begin{itemize}
        \item Type: int
        \item Default: \texttt{-1}
    \end{itemize}
    
    \item \textbf{\texttt{inactive\_mark}}: (Optional) Marked as deprecated or pending removal.
    \begin{itemize}
        \item Type: bool
        \item Default: \texttt{False}
    \end{itemize}
\end{itemize}

\subsubsection{Statistical Fields}

These fields are used to extend knowledge metadata and statistics.

\begin{itemize}
    \item \textbf{\texttt{metadata}}: (Optional) Extended metadata for storing additional, non-standardized information.
    \begin{itemize}
        \item Type: Dict[str, Any]
        \item Default: empty dict \texttt{\{\}}
    \end{itemize}
    
    \item \textbf{\texttt{profile}}: (Optional) Usage statistics of the knowledge.
    \begin{itemize}
        \item Type: Dict[str, Any]
        \item Default: empty dict \texttt{\{\}}
    \end{itemize}
\end{itemize}

\subsubsection{Internal Fields}

These fields are used internally by the system for identification, content tracking, and optimization.

\begin{itemize}
    \item \textbf{\texttt{\_id}}: Unique identifier for the knowledge, computed from key identity fields. If not set, auto-generated from \texttt{identity\_hash\_fields}.
    \begin{itemize}
        \item Type: Optional[str]
        \item Default: \texttt{None}
        \item \texttt{identity\_hash\_fields = ["type", "name", "version", "variant", "source", "creator", "owner", "workspace", "collection", "tags", "timefluid"]}
    \end{itemize}
    
    \item \textbf{\texttt{\_content\_hash}}: Encrypted hash of the content, used to detect content changes. If not set, auto-generated from \texttt{content\_hash\_fields}.
    \begin{itemize}
        \item Type: Optional[str]
        \item Default: \texttt{None}
        \item \texttt{content\_hash\_fields = ["content", "content\_resources"]}
    \end{itemize}
    
    \item \textbf{\texttt{\_slots}}: Slot cache for tag classification, computed from \texttt{tags}.
    \begin{itemize}
        \item Type: Dict[str, Set[str]]
        \item Default: empty dict \texttt{\{\}}
    \end{itemize}
\end{itemize}

\end{sloppypar}

\subsection{\ukf Template Example}

To define a new \ukf template, create a Python class that inherits from the \texttt{BaseUKF} class. Typically, it is sufficient to override the type and content composers. Consider the \texttt{ColumnRule} example:
\begin{lstlisting}[language=python, caption={Definition of \texttt{ColumnRule} by inheriting \texttt{BaseUKF}.}, label={code:python-ukf}]
def format_column_rule(kl, **kwargs):
    column_prompt = load_txt("& prompts/rules/column.txt")
    synonyms = kwargs.get('matches', unique([kl.name]+kl.synonyms))
    table_id = kwargs.get('table_id', kl.content_resources.get('table_id'))
    physical = kl.content_resources.get('predicate', dict()).get('physical')
    return column_prompt.format(
        synonyms = "/".join(set(f'"{syn}"' for syn in synonyms)),
        table_id = table_id,
        column_id = physical,
        comment = kl.name
    )
class ColumnRule(BaseUKF):
    type: str = Field(default='column', description="Knowledge type identifier", frozen=True)
    content_composers: Dict[str, Callable] = Field(default={"default": format_column_rule}, description="Content processing functions", frozen=True)
\end{lstlisting}

Note that the content composer's prompt is stored externally in \texttt{prompts/rules/column.txt} to facilitate internationalization. For instance, an English \texttt{ColumnRule} composer prompt might be:
\begin{lstlisting}[caption={Example \texttt{ColumnRule} composer prompt.}, label={code:prompt-column}]
- {synonyms} in query could be referring to column `"{table_id}"."{column_id}" -- Column: {comment}`.
\end{lstlisting}

After defining the content composer, based on \texttt{table\_id}, \texttt{column\_id}, and \texttt{comment}, the knowledge can be dynamically serialized upon use. The external parameter \texttt{matches}, generated during retrieval, links the original query phrases to this specific knowledge.

\section{SQL Profile Example}
\label{sec:appendix-sql}
\begin{sloppypar}

\begin{lstlisting}[language=SQL, caption={Example of a profiled SQL (PostgreSQL).}, label={code:sql-example2}]
-- User Query: Show me the top 20 enterprise parent customers in ME Outlets based on their YoY admin expenses for August through December 2023.
-- User Profile:
--   - Default Currency: USD
--   - Default Caliber: B
-- Query Time: 2025-05-01
-- Expected SQL Result Schema: ("enterprise_parent_customer", "2022-08 to 2022-12", "2023-08 to 2023-12", "YOY Growth")
-- Key steps/corner-cases/knowledge/formula:
-- 1. Date Range Handling: For a range like 'YYYY-MM' to 'YYYY-MM', use the start and end months ('YYYYMM') in the WHERE clause's period filter.
-- 2. YTD Calculation for a Range: Sum the YTD amount at the end of the period and subtract the YTD amount at the start of the period (minus one month). This is achieved by summing `ytd_amount` for the end month and `-ytd_amount` for the start month. Do the same for PY_YTD amounts.
-- 3. YoY Growth: Calculated as (Current Period Amount - Prior Period Amount) / ABS(Prior Period Amount). Handle division by zero by returning 0.0.
-- 4. Filters: Apply filters for `dealer_lv4` based on "ME_Outlets" from example, and `report_l2` for "Administrative Expenses".
-- 5. Currency and Caliber: Use `usd_ytd_amt_b` and `usd_py_ytd_amt_b` as specified.
-- 6. Top N: Order by the calculated "YOY Growth" in descending order and apply LIMIT 20.
SELECT
    enterprise_parent_customer,
    -- Calculate Prior Year Period Sum (2022-08 to 2022-12)
    COALESCE(SUM(CASE
        WHEN period = '202312' THEN usd_py_ytd_amt_b -- YTD value at the end of the period for prior year
        WHEN period = '202307' THEN -usd_py_ytd_amt_b -- YTD value at the beginning of the period (subtracted to get the range sum)
        ELSE 0
    END), 0) AS "2022-08 to 2022-12",
    -- Calculate Current Year Period Sum (2023-08 to 2023-12)
    COALESCE(SUM(CASE
        WHEN period = '202312' THEN usd_ytd_amt_b -- YTD value at the end of the period for current year
        WHEN period = '202307' THEN -usd_ytd_amt_b -- YTD value at the beginning of the period (subtracted to get the range sum)
        ELSE 0
    END), 0) AS "2023-08 to 2023-12",
    -- Calculate Year-over-Year Growth
    CASE
        -- Alias for Prior Year Sum to avoid repetition and improve readability
        WHEN (COALESCE(SUM(CASE
            WHEN period = '202312' THEN usd_py_ytd_amt_b
            WHEN period = '202307' THEN -usd_py_ytd_amt_b
            ELSE 0
        END), 0)) = 0 THEN 0.0
        ELSE (
            (COALESCE(SUM(CASE
                WHEN period = '202312' THEN usd_ytd_amt_b
                WHEN period = '202307' THEN -usd_ytd_amt_b
                ELSE 0
            END), 0)) -- Current Year Sum
            -
            (COALESCE(SUM(CASE
                WHEN period = '202312' THEN usd_py_ytd_amt_b
                WHEN period = '202307' THEN -usd_py_ytd_amt_b
                ELSE 0
            END), 0)) -- Prior Year Sum
        ) / ABS(
            COALESCE(SUM(CASE
                WHEN period = '202312' THEN usd_py_ytd_amt_b
                WHEN period = '202307' THEN -usd_py_ytd_amt_b
                ELSE 0
            END), 0)
        )
    END AS "YOY Growth"
FROM
    xxaw_fi.revrec_sum_v
WHERE
    -- Filter for the specific period (start and end months of the range)
    period IN ('202307', '202312')
    AND
    -- Filter for "admin expenses" as inferred from example
    report_l2 = 'Administrative Expenses'
    AND
    -- Filter for "ME Outlets" as inferred from example
    dealer_lv4 IN ('Rivendell Rapid Service', 'Shire Auto Outlet')
GROUP BY
    enterprise_parent_customer
ORDER BY
    "YOY Growth" DESC
LIMIT 20;
\end{lstlisting}
\end{sloppypar}

\section{\benchI Example}
\label{sec:appendix-bench}
\begin{sloppypar}

\begin{lstlisting}[caption={Example \benchI inputs (serialized): question with user profile and temporal context.}, label={code:nl-example1}]
Question: Where does Nova rank for Inventory Impairment Loss BCR in Sweden and Spain this month?
User Profile:
- Default Currency: EUR
- Default Caliber: B
- Default Region: region=Western Europe >> country=Germany >> national_area=Bavaria
- Default Department: vehicle_brand_group=Nova Dynamics
Query Time: 2023-12-21
\end{lstlisting}

\begin{lstlisting}[language=SQL, caption={Example \benchI output SQL (PostgreSQL).}, label={code:sql-example1}]
SELECT
  *
FROM (
  SELECT
    currency,
    CASE
      WHEN SUM(CASE WHEN period BETWEEN '202311' AND '202312' THEN ptd_budget_amt ELSE 0 END) = 0
      THEN 0.0
      ELSE SUM(ptd_amt) / SUM(CASE WHEN period BETWEEN '202311' AND '202312' THEN ptd_budget_amt ELSE 0 END)
    END AS "bcr",
    RANK() OVER (
      ORDER BY CASE
        WHEN SUM(CASE WHEN period BETWEEN '202311' AND '202312' THEN ptd_budget_amt ELSE 0 END) = 0
        THEN 0.0
        ELSE SUM(ptd_amt) / SUM(CASE WHEN period BETWEEN '202311' AND '202312' THEN ptd_budget_amt ELSE 0 END)
      END DESC
    ) AS rank
  FROM xxaw_fi.bud_fcst_actual_v
  WHERE
    (
      country IN ('Sweden', 'Spain')
      AND vehicle_brand_group = 'Nova Dynamics'
      AND period BETWEEN '202311' AND '202312'
      AND report_l3 IN ('Inventory Impairment Loss')
      AND caliber IN ('B')
    )
  GROUP BY
    currency
  ORDER BY
    bcr DESC
) AS subquery_alias
WHERE
  currency IN ('EUR')
\end{lstlisting}

\end{sloppypar}




\balance